\documentclass[traditabstract]{aa} 

\newcommand{\gtsim}{\protect\raisebox{-0.5ex}{$\:\stackrel{\textstyle >}
        {\sim}\:$}}
\newcommand{\ltsim}{\protect\raisebox{-0.5ex}{$\:\stackrel{\textstyle <}
        {\sim}\:$}}

\usepackage{longtable}    
\usepackage{aalongtable}    
\usepackage{latexsym}
\usepackage{graphicx}
\usepackage{epsfig}
\usepackage{lscape}
\usepackage[authoryear]{natbib}
\bibpunct{(}{)}{;}{a}{}{,} 

\begin{document}

\title{On the accretion properties of young stellar objects in the
L1615/L1616 cometary cloud\thanks{Based on FLAMES (UVES+GIRAFFE) observations collected 
at the Very Large Telescope (VLT; Paranal, Chile). Program 076.C-0385(A).} } 
   
\author{K. Biazzo \inst{1} \and J. M. Alcal\'{a}\inst{2} \and A. Frasca \inst{1} \and M. Zusi\inst{2} 
       \and F. Getman \inst{2} \and  E. Covino\inst{2} \and D. Gandolfi\inst{3}}

\offprints{K. Biazzo}
\mail{katia.biazzo@oact.inaf.it}

\institute{INAF - Osservatorio Astrofisico di Catania, via S. Sofia, 78, I-95123 Catania, Italy
\and INAF - Osservatorio Astronomico di Capodimonte, via Moiariello, 16, I-80131 Napoli, Italy
\and Landessternwarte K\"onigstuhl, Zentrum f\"ur Astronomie der Universit\"at, Heidelberg, K\"onigstuhl 12, D-69117 Heidelberg, Germany
}

\date{Received 21 May 2014/ accepted 07 September 2014}

\abstract{We present the results of FLAMES/UVES and FLAMES/GIRAFFE spectroscopic observations of 23 low-mass 
stars in the L1615/L1616 cometary cloud, complemented with FORS2 and VIMOS spectroscopy of 31 additional stars in the 
same cloud. L1615/L1616 is a cometary cloud where the star formation was triggered by the impact of the massive stars in the 
Orion OB association. From the measurements of the lithium abundance and radial velocity, we confirm the membership of 
our sample to the cloud. We use the equivalent widths of the H$\alpha$, H$\beta$, and the \ion{He}{i} $\lambda$5876, 
$\lambda$6678, $\lambda$7065 \AA~emission lines to calculate the accretion luminosities, $L_{\rm acc}$, and the mass 
accretion rates, $\dot M_{\rm acc}$. We find in L1615/L1616 a fraction of accreting objects ($\sim 30\%$), which is 
consistent with the typical fraction of accretors in T associations of similar age ($\sim 3$~Myr). The mass accretion rate 
for these stars shows a trend with the mass of the central object similar to that found for other star-forming regions, with 
a spread at a given mass which depends on the evolutionary model used to derive the stellar mass. Moreover, the behavior 
of the $2MASS/WISE$ colors with $\dot M_{\rm acc}$ indicates that strong accretors with $\log \dot M_{\rm acc} \gtsim -8.5$ 
dex show large excesses in the $JHK{\rm s}$ bands, as in previous studies. We also conclude that the accretion properties of 
the L1615/L1616 members are similar to those of young stellar objects in T associations, like Lupus.}
   
\maketitle
	   
\titlerunning{Accretion properties in young low-mass stars in L1615/L1616}
\authorrunning{K. Biazzo et al.}

\section{Introduction}

The Lynds 1616 cloud (hereafter L1616; \citealt{lynds1962}), at a distance of about 
450 pc, forms, together with Lynds 1615 (hereafter L1615), a cometary-shaped cloud west of the 
Orion OB association  ($\alpha \sim 5^{\rm h} 7^{\rm m}$, $\delta \sim -3{\degr} 20'$; see review 
by \citealt{alcalaetal2008}, and references therein). It extends about $40'$ in the sky and 
shows evidence of  ongoing star formation activity that might have been triggered 
by the ultraviolet (UV) radiation coming from the massive stars in the Orion OB association 
(see \citealt{stankeetal2002}, and references therein). In particular, recent studies 
led by \cite{leechen2007} support the validity of the radiation-driven implosion mechanism, where 
the UV photons from luminous massive stars create expanding ionization fronts to evaporate and compress 
nearby clouds into bright-rimmed or comet-shaped clouds, like L1615/L1616. Implosive pressure 
then causes dense clumps to collapse, prompting the formation of stars. Young stars in comet-shaped 
clouds are therefore likely to have been formed by a triggering mechanism.

\cite{alcalaetal2004} reported a sample of 33 young stellar objects (YSOs) associated  
with L1615/L1616, while \cite{gandolfietal2008} performed a comprehensive census of 
the pre-main sequence (PMS) population in L1615/L1616, which consists of 56 YSOs. 
These two works were focused on the investigation of the star formation history, 
the relevance of the triggered scenario, and the initial mass function, but no 
study on accretion properties was addressed. As a continuation of these works, here we use 
further spectroscopic observations to derive the accretion luminosity, $L_{\rm acc}$, 
and the mass accretion rate, $\dot M_{\rm acc}$, of a sample of low-mass YSOs in L1615/L1616. 
We also investigate whether the accretion properties of young stellar 
objects in a cometary cloud like L1615/L1616 are similar to those of PMS stars in T associations, 
like Lupus, Taurus or Chamaeleon.

The outline of the paper is as follows. In Sect.~\ref{sec:obs}, we describe the 
spectroscopic observations, the data reduction, and the sample investigated. 
In Sect.~\ref{sec:accretion_rates}, several accretion diagnostics are used 
to derive the mass accretion rates.
The main results on the accretion and infrared (IR) properties are discussed 
in Sect.~\ref{sec:discussion}, while our conclusions are presented 
in Sect.~\ref{sec:conclusions}\footnote{Three appendixes present additional 
material on: radial velocity and lithium abundance measurements, comparison between 
$\dot M_{\rm acc}$ derived through three different PMS evolutionary tracks, and 
notes on individual objects.}. 

\section{Observations, data reduction, and the sample}
\label{sec:obs}

\subsection{FLAMES observations and data reduction}
The observations were conducted in February-March 2006 in visitor mode using FLAMES (UVES+GIRAFFE) at the
VLT. The CD\#3 cross-disperser and the LR6 grating were used for the UVES and GIRAFFE 
spectrographs, respectively. A brief summary of the observations is given in 
Table~\ref{tab:logs}, while the complete journal of the observations is given in
Table~\ref{tab:all_param}. We observed 23 low-mass ($0.1 \ltsim M_\star \ltsim 2.3 M_\odot$) 
objects with GIRAFFE in the MEDUSA
mode\footnote{This is the observing  mode of FLAMES in which 132 fibers each with a
projected diameter on the sky of 1\farcs{2}, feed the GIRAFFE spectrograph. Some
fibers are set on the target stars and others on the sky background.}; one target 
(the classical T Tauri star - CTTs - LkH$\alpha$~333) was observed with both spectrographs. 
Nineteen objects were observed several (2--7) times within 2 days 
(see Table~\ref{tab:all_param}).  

The reduction of the UVES spectra was performed using the pipeline developed by 
\cite{modiglianietal2004}, which includes the following steps: subtraction of a 
master bias, \'echelle order definition, extraction of thorium-argon spectra, 
normalization of a master flat-field, frame extraction, wavelength calibration, 
and correction of the science frame for the normalized master flat-field. 
Sky subtraction was performed with the IRAF\footnote{IRAF is distributed by the
National Optical  Astronomy Observatory, which is operated by the Association of 
the Universities for Research in Astronomy, inc. (AURA) under cooperative agreement 
with the National Science Foundation.} task {\sc sarith} using the fibers 
allocated to the sky.

The GIRAFFE data were reduced using the GIRAFFE Base-Line Data Reduction Software
1.13.1 (girBLDRS; \citealt{blechaetal2000}): bias and flat-field subtraction, correction
for the fiber transmission coefficient, wavelength calibration, and science frame
extraction were  performed. Then, a sky correction was applied to each stellar
spectrum using the task {\sc sarith} in the IRAF {\sc echelle} package and by subtracting the
average of several sky spectra obtained simultaneously during a given night. 

\setlength{\tabcolsep}{2.5pt}
\begin{table}
\caption{Summary of the observations.} 
\label{tab:logs}
\begin{center}
\begin{tabular}{lccccc}
\hline
\hline
Instrument &   Range & Resolution & \# stars & \# spectra\\ 
	   &   (\AA) & ($\lambda/\Delta\lambda$) &  & \\ 
\hline
UVES	& 4764--6820 & 47\,000 & 1 & 6 \\
GIRAFFE & 6438--7184 &  8\,600 & 23 & 53 \\
\hline
\end{tabular}
\end{center}
\end{table}
\normalsize

\subsection{The sample}
\label{sec:sample}
Since our goal is to investigate the accretion and the IR properties of the YSOs 
in the cometary cloud, we need a well characterized sample of YSOs 
both in terms of their physical parameters and their association with the cloud, 
as well as in terms of their accretion diagnostics and IR colors. 

The stellar parameters (spectral types, effective temperatures, luminosities, and masses) 
were derived by \cite{gandolfietal2008}. We adopt those determinations here. We note that one 
object, namely TTS~050730.9$-$031846, has a significantly lower luminosity compared to the other 
objects in the sample (see Fig.~3 in \citealt{gandolfietal2008}). This most-probable sub-luminous 
object is further discussed in Appendix~\ref{sec:notes_on_ind_objects}. 
 
Regarding the association with the cloud, we investigated the kinematics by means of radial velocity 
(RV) determinations, $V_{\rm rad}$, and lithium abundance, $\log n{\rm (Li)}$, following the same methods 
as in \cite{biazzoetal2012}. The details of such determinations can be found in Appendix~\ref{sec:rad_vel}. 
The radial velocity distribution of the YSOs in L1615/L1616 has an average of 
$\langle V_{\rm rad}\rangle=23.2 \pm 3.1$ km s$^{-1}$ (see Fig.~\ref{fig:vrad_distr}), which is consistent with 
that reported by \cite{alcalaetal2004} (i.e. $\langle V_{\rm rad}\rangle=22.3\pm 4.6$ km s$^{-1}$), 
and in general with that of the Orion complex (see, e.g., \citealt{briceno2008, biazzoetal2009, sergisonetal2013}). 
Likewise, the average lithium abundance is $\log n{\rm (Li)}$ $\sim 3.3$\,dex with a 
dispersion of $\pm 0.3$ dex (see Table \ref{tab:all_param} and Appendix \ref{sec:lithium_abundance}). 
Both radial velocities and lithium abundances confirm that all targets studied in this work 
are associated with the cometary cloud.

\begin{figure}[t!]
\begin{center}
 \begin{tabular}{c}
\hspace{-.6cm}
\includegraphics[width=9.5cm]{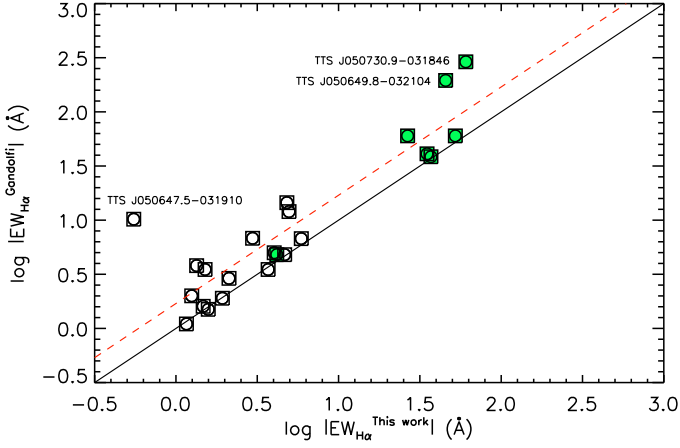}
\vspace{-.4cm}
 \end{tabular}
\caption{Comparison between our  $EW_{\rm H\alpha}$ values and those obtained by 
\cite{gandolfietal2008}. The solid line represents the 1:1 relation, while the dashed 
one is shifted by the r.m.s. difference between the $\log EW_{\rm H\alpha}$ values 
(excluding the three most deviating objects labeled in the figure). Filled symbols 
refer to most-probable accreting stars (see text and Fig.~\ref{fig:2MASS_color_color}).}
\label{fig:ew_comparison}
 \end{center}
\end{figure}

In order to have a more complete sample, we included in our analysis the YSOs lacking FLAMES spectroscopy, 
but for which \cite{gandolfietal2008} have provided measurements of the H$\alpha$ equivalent width from 
FORS2@VLT and VIMOS@VLT low-resolution spectra acquired in February-March 2003. Figure~\ref{fig:ew_comparison} 
shows the comparison between our H$\alpha$ equivalent widths\footnote{Equivalent widths of all lines used 
in the present work as accretor diagnostics were measured by direct integration using the IRAF task {\sc splot}. 
As errors in the line equivalent widths, the standard deviations of three measurements were adopted.} ($EW_{\rm H\alpha}$) 
and the measurements reported by \cite{gandolfietal2008} for the stars in common. Although there is a
general agreement, the \cite{gandolfietal2008} $EW_{\rm H\alpha}$ are systematically higher 
than ours in average by about 4 \AA~(excluding the three most deviating stars). We believe that 
this systematic difference is due to the lower spectral resolution used by \cite{gandolfietal2008} 
with respect to the resolution of our FLAMES spectra, whereas for the three YSOs that deviate 
significantly from the 1:1 relationship in Fig.~\ref{fig:ew_comparison} the differences are most likely 
related to variability. Two of these objects will result to be accretors (see later on) and their variability 
will be discussed in Sections \ref{sec:sub-luminous} and \ref{sec:TTS050649.8-032104}, while the 
other is a weak lined T~Tauri star (WTTs) and the difference of $\sim 10$~\AA~between our $EW_{\rm H\alpha}$ and 
the values by \cite{gandolfietal2008} could be related to stellar activity phenomena. This comparison 
justifies in the following analysis the use of the \cite{gandolfietal2008} $EW_{\rm H\alpha}$ values 
for the stars not observed by us. Therefore, our analysis is based on 23 objects observed by us with FLAMES, 
and 31 targets previously observed by \cite{gandolfietal2008} at low resolution. All these objects are listed 
in Tables~\ref{tab:2mass_wise} and \ref{tab:ew_flux}. 

We used the criteria of \cite{whitebasri2003} based on 
spectral types and $EW_{\rm H\alpha}$ to distinguish between accretors and non-accretors. 
In this way, a total of 15 YSOs in L1615/L1616 can be classified as accretors, 7 within our 
sample and 8 within the \cite{gandolfietal2008} sample. Note that TTS\,050649.8$-$031933, 
originally classified as a WTTs by \cite{gandolfietal2008} is tagged here as accretor 
because its spectrum shows helium and forbidden oxygen lines in emission (see 
Sect.~\ref{TTS050649.8-031933}), typical of accreting objects (see their Table 4).

In addition, we considered ancillary IR data both from the Two-Micron All Sky Survey ($2MASS$; 
\citealt{cutrietal2003}) and from the Wide-field Infrared Survey Explorer ($WISE$; 
\citealt{cutrietal2012}) catalogues (see Table~\ref{tab:2mass_wise}) to investigate the IR 
properties of the sample and their possible correlation with accretion diagnostics. 

\begin{figure}	
\begin{center}
 \begin{tabular}{c}
\includegraphics[width=9.2cm]{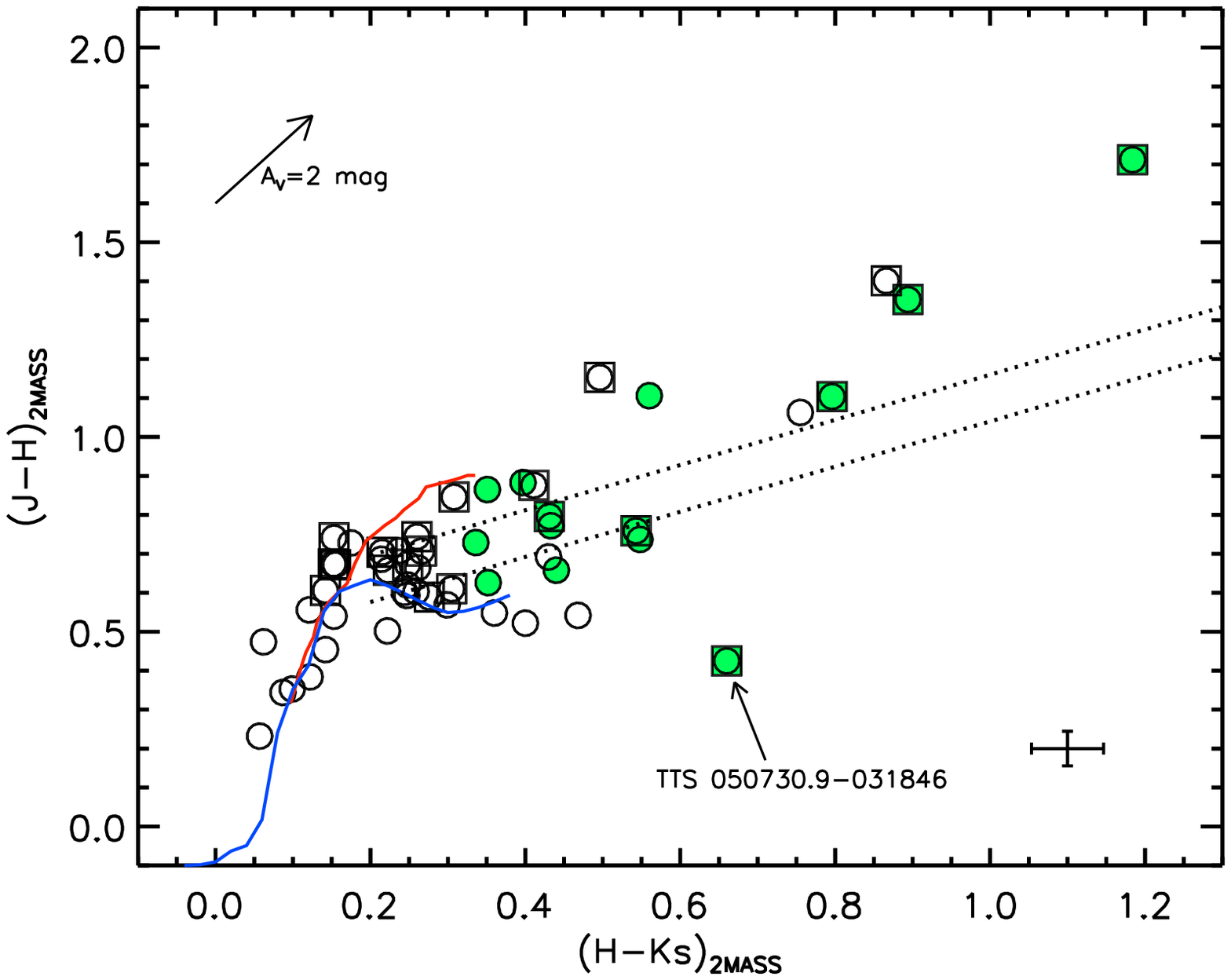}
\vspace{-.2cm}
 \end{tabular}
\caption{{\it 2MASS} color-color diagram of the L1615/L1616 targets. Open circles represent 
low-mass stars analyzed by \cite{gandolfietal2008}, while squares mark the targets observed 
for this work with the FLAMES spectrograph. Filled symbols are the most probable accretors 
as defined in Section~\ref{sec:sample}. The dwarf (lower branch; \citealt{bessellbrett1988}) 
and the giant (upper branch; \citealt{kenyonhartmann1995}) sequences are 
marked by solid lines. The arrow at the top upper-left corner indicates the reddening vector 
for $A_{V}=2$~mag. The CTTs locus (\citealt{meyeretal1997}) is delimited by the dotted lines. 
The position of the sub-luminous candidate TTS~050730.9$-$031846 is also displayed. 
The mean {\it 2MASS} photometric errors are overplotted on the lower-right corner of 
the panel.}
\label{fig:2MASS_color_color}
 \end{center}
\end{figure}

\begin{figure*}	
\begin{center}
 \begin{tabular}{c}
\hspace{-.6cm}
\includegraphics[width=6.7cm]{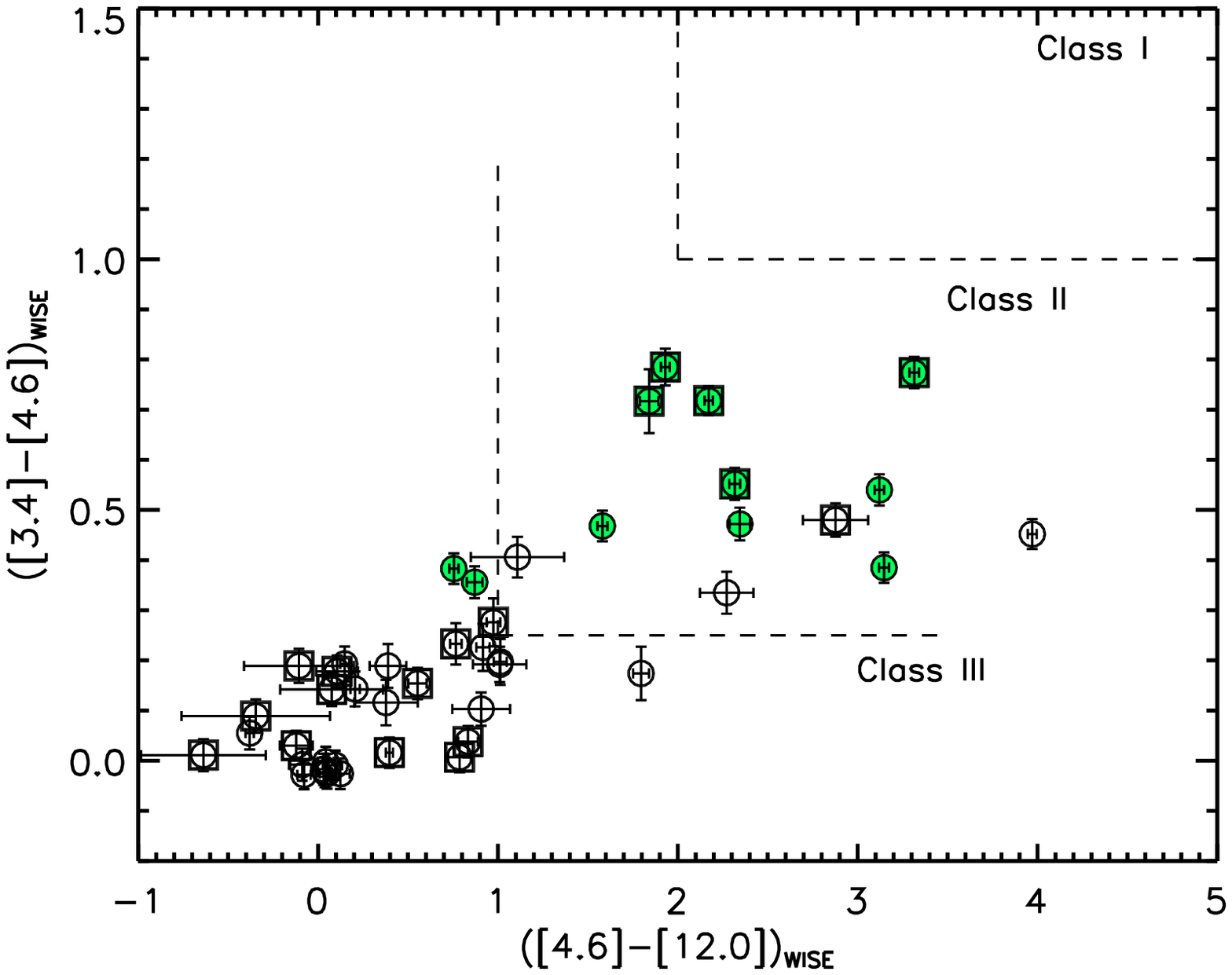}
\hspace{-.7cm}
\includegraphics[width=6.7cm]{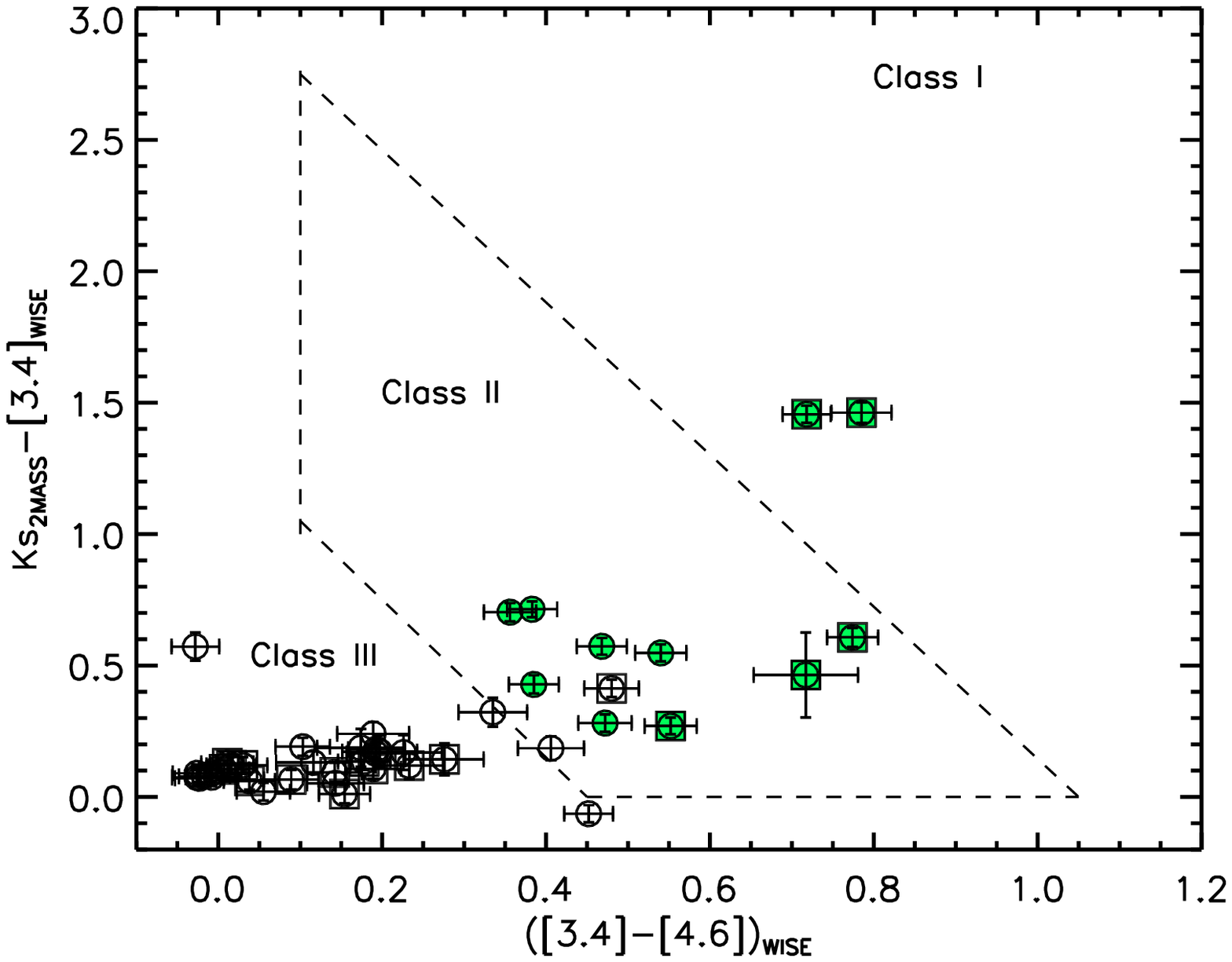}
\hspace{-.7cm}
\includegraphics[width=6.7cm]{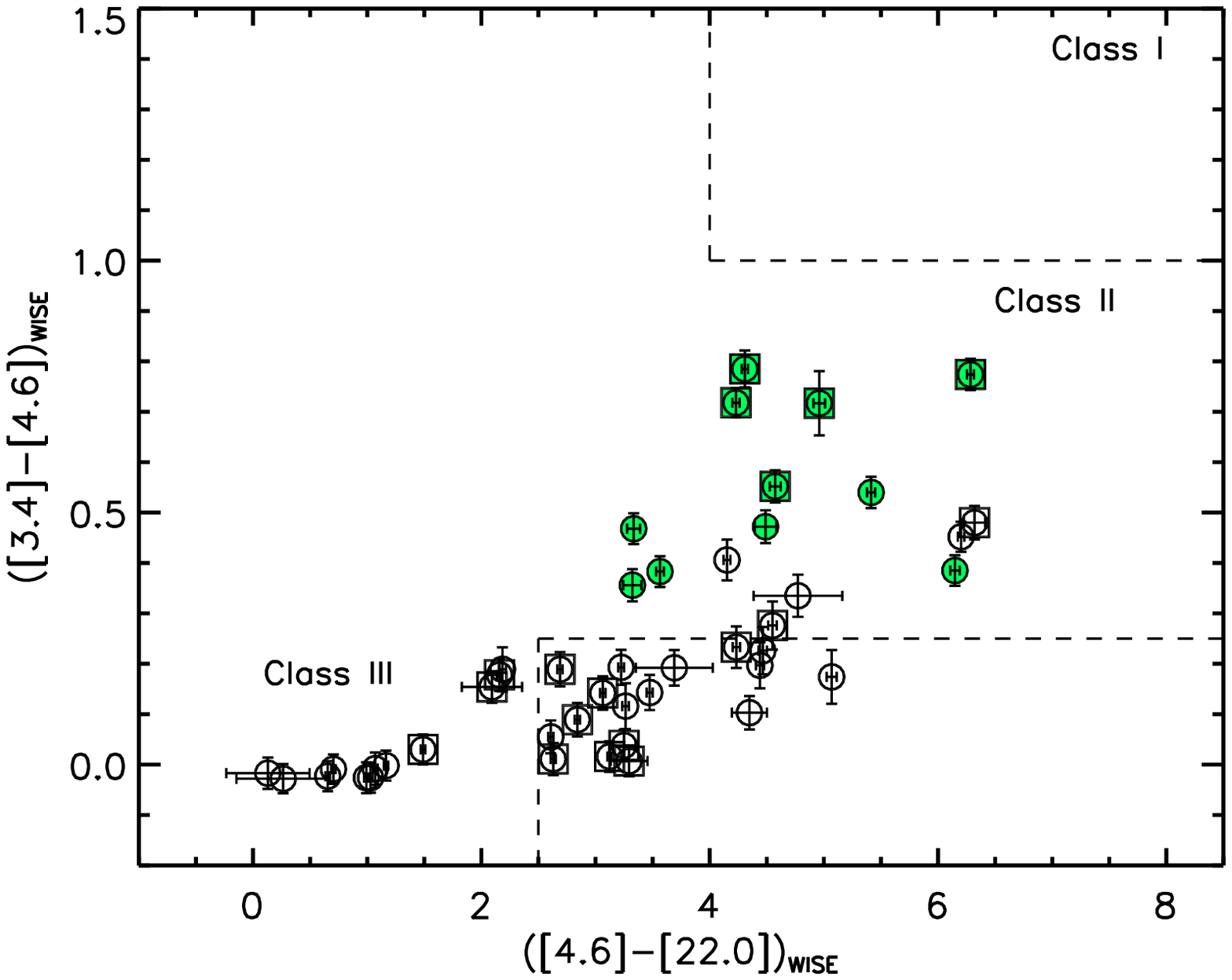}
\vspace{-.2cm}
 \end{tabular}
\caption{{\it WISE}/{\it 2MASS} color-color diagrams of the L1615/L1616 targets. 
Dashed lines indicate the boundaries of Class I, Class II, and disk-less stars 
(Class III objects) as defined in \cite{koenigetal2012}. 
The non-accretors falling within the Class II region  will be discussed in 
Appendix~\ref{sec:notes_on_ind_objects}. Symbols are as in Fig.~\ref{fig:2MASS_color_color}. 
}
\label{fig:WISE_color_color}
 \end{center}
\end{figure*}

The position in the $2MASS$ color-color diagram of all YSOs classified here as accretors 
and non-accretors is shown in Fig.~\ref{fig:2MASS_color_color} (filled and opened symbols, respectively). 
All YSOs classified by us as accretors show near-IR excess. The anomalous colors 
of TTS~050730.9$-$031846 are consistent with its sub-luminous nature (see Section~\ref{sec:sub-luminous}). 
The three non-accretors with apparently infrared excess represent stars with high 
values of $A_{V}$, as reported by \cite{gandolfietal2008}. Their $WISE$ colors are typical 
of Class~III objects, confirming their WTTs nature (see Fig.~\ref{fig:WISE_color_color}). 
Moreover, note that all YSOs classified here as accretors have $WISE$ colors typical of 
Class~II YSOs. 

Summarizing, we find a fraction of accretors in L1615/L1616 ($\sim$30\%) consistent 
within the errors with the fraction of disks recently reported by \cite{ribasetal2014} for an average 
age of 3\,Myr (see their Fig.~2).

\section{Accretion diagnostics and mass accretion rates}
\label{sec:accretion_rates}

According to the magnetospheric accretion model (\citealt{uchida85, konigl91, shu94}), 
matter is accreted from the disk onto the star and shocks the stellar surface producing high temperature  
($\sim 10^4$ K) gas, giving rise to emission in the blue continuum and in many lines, 
which can be observed as photometric and spectroscopic diagnostics. Primary accretion 
diagnostics, such as the UV excess emission, the Paschen/Balmer continua, and 
the Balmer jump (see, e.g., \citealt{herczeghillenbrand2008, alcalaetal2014}), and 
secondary tracers, like hydrogen recombination lines and the \ion{He}{i}, \ion{Ca}{ii}, \ion{Na}{i} 
lines (see, e.g., \citealt{muzerolleetal1998, antoniuccietal2011, biazzoetal2012}) are therefore 
useful tools to detect accretion signatures and to derive the energy losses due 
to accretion, i.e. the accretion luminosity (e.g., \citealt{gullbringetal1998, 
herczeghillenbrand2008, rigliacoetal2011b, ingleby2013, alcalaetal2014}).

In the context of the magnetospheric accreting model, the accretion luminosity can be 
converted into mass accretion rate, $\dot M_{\rm acc}$, using the following relationship 
\citep{hartmann1998}:

\begin{equation}
\label{for:Macc}
\dot{M}_{\rm acc} = \left(1 - \frac{R_{\star}}{R_{\rm in}} \right)^{-1} \frac{L_{\rm acc} R_{\star}}{G M_{\star}}\,
\approx 1.25 \frac{L_{\rm acc} R_{\star}}{G M_{\star}}\,,
\end{equation} 
 
\noindent{where $M_\star$ and $R_\star$ are the stellar mass and radius, respectively,
$R_{\rm in}$ is the YSO inner-disk radius, and $G$ is the universal gravitational constant. 
$R_{\rm in}$ corresponds to the distance at which the disk is truncated -- 
due to the stellar magnetosphere -- and from which the disk gas is accreted, 
channeled by the magnetic field lines. In previous works, it has been assumed 
that $R_{\rm in}$ is $\sim 5\,R_\star$ (see, e.g., \citealt{alcalaetal2011}). }

The accretion luminosity can be estimated from empirical linear relationships
between the observed line luminosity, $L^{\lambda}$, and $L_{\rm acc}$ derived through 
primary diagnostics (see, e.g., \citealt{gullbringetal1998, herczeghillenbrand2008, alcalaetal2014}).
Such relationships have been established by the simultaneous observations of 
many accretion indicators and by modeling the continuum excess emission.

For the accretors in our sample, we used the luminosity of several emission lines 
(H$\alpha$ $\lambda$6563 \AA, H$\beta$ $\lambda$4861 \AA, \ion{He}{i} $\lambda$5876
\AA, \ion{He}{i} $\lambda$6678 \AA, and \ion{He}{i} $\lambda$7065 \AA) within the wavelength 
range covered by the FLAMES spectra, while for the objects in \cite{gandolfietal2008} we 
used the H$\alpha$ line. We then considered the recent $L^{\lambda}-L_{\rm acc}$ relations 
by \cite{alcalaetal2014} to derive the accretion luminosity from each line. These relationships 
consider a combination of all accretion indicators calibrated on sources for which the UV excess 
emission and the Paschen/Balmer continua were measured simultaneously.

Unfortunately, we do not have simultaneous or quasi-simultaneous photometry in hand and 
our fiber-fed spectra cannot be calibrated in flux. Therefore, the best approach to calculate line 
luminosities from our data is by deriving line surface fluxes using the equivalent widths and 
assuming continuum fluxes from model atmospheres. Thus, the line luminosity $L^{\lambda}$ was calculated 
using the same approach as in \cite{biazzoetal2012}. In particular, $L^{\lambda} = 4 \pi R_\star^2 F^{\lambda}$, 
where the stellar radius was taken from \cite{gandolfietal2008} and the 
surface flux, $F^{\lambda}$, was derived by multiplying the EW of each line 
($EW_{\lambda}$) by the continuum flux at wavelengths adjacent to the line 
($F_{\rm cont}^{\lambda\pm\Delta\lambda}$). The latter was gathered from the 
NextGen Model Atmospheres (\citealt{hauschildtetal1999}), assuming the corresponding 
YSO effective temperature and surface gravity. The gravity was estimated
for every YSO from the mass and stellar radius reported in \cite{gandolfietal2008}. 
In particular, we considered the three different $M_\star$ values provided by the authors 
for three different sets of PMS evolutionary 
tracks (namely, \citealt{baraffeetal1998} and \citealt{chabrieretal2000}, 
\citealt{dantonamazzitelli1997}, \citealt{pallastahler1999}; hereafter Ba98+Ch00, 
DM97, PS99, respectively, as used in \citealt{gandolfietal2008}). We stress that the 
mean difference in $\log g$ coming from the use of the three evolutionary tracks 
varies from $\sim$0.0 to 0.4 dex. Such a kind of differences in $\log g$ may 
lead to an uncertainty in the continuum flux of less than $\sim 10\%$, mainly depending on the 
effective temperature, the surface gravity itself, and the line considered. This represents 
the typical error in the continuum flux we considered for the estimation of the uncertainty in 
$\dot M_{\rm acc}$ (see text later on).

In the end, the different line diagnostics considered by us yielded consistent 
values of $L_{\rm acc}$, which justified the use of all of them to compute an 
average $\langle L_{\rm acc}\rangle$ for each YSO (see Table~\ref{tab:accret_param}).
In this way, the error on the average $L_{\rm acc}$ derived from several 
diagnostics, measured simultaneously, is minimized, as found by \cite{rigliacoetal2012} 
and \cite{alcalaetal2014}. The mass accretion rate ($\dot M_{\rm acc}$) was then calculated using 
$\langle L_{\rm acc}\rangle$ and the Eq.~\ref{for:Macc}, and adopting the 
$M_\star$ and $R_\star$ values reported in \cite{gandolfietal2008}. For every accretor, 
we thus derived three values of $\dot M_{\rm acc}$ using the three values of $M_\star$
(Table~\ref{tab:accret_param}).

Contributions to the error budget on $\dot M_{\rm acc}$ include uncertainties 
on stellar mass, stellar radius, inner-disk radius, and $L_{\rm acc}$. 
Assuming mean errors of $\sim 0.1 M_\odot$ in $M_\star$ and $\sim 0.1 R_\odot$ in 
$R_\star$ (\citealt{gandolfietal2008}), $1-13\%$ as relative error in $EW_{\lambda}$, 
10\% in $F_{\rm cont}^{\lambda\pm\Delta\lambda}$, and the uncertainties in 
the relationships by \cite{alcalaetal2014}, we estimate a typical error in 
$\log \dot M_{\rm acc}$ of $\sim 0.5$ dex. 

Note that the equivalent width values are not corrected for veiling, which alters 
the continuum of the spectra. In case of strong accretors, 
the continuum excess emission becomes important, but we quantify this effect on our sample 
in the next section.

\subsection{Impact of veiling on the $\dot M_{\rm acc}$ estimates}
\label{sec:veiling}

We estimate how the amount of veiling affects the $\dot M_{\rm acc}$ estimates by 
running the ROTFIT\footnote{ROTFIT is an IDL code. IDL (Interactive Data Language) is a registered 
trademark of Exelis Visual Information Solutions.} code (\citealt{frascaetal2003,frascaetal2006}) on the 
spectra of our accretors. This code compares the target spectrum with a grid of slowly-rotating 
and low-activity templates, aligned with the target spectrum, re-sampled on its spectral 
points, and artificially broadened with a rotational profile until the minimum of the residuals 
is reached (see details in Frasca et al. 2014, submitted). In order to find the best template 
reproducing the veiled accretors, we included the veiling as an additional parameter. 
This was done by adding a featureless veiling to each template, whose continuum-normalized 
spectrum becomes: 
\begin{equation}
\label{for:veiling}
\Biggl(\frac{F_{\lambda}}{F_{\rm cont}}\Biggr)^{veil}=\frac{\Bigl(\frac{F_{\lambda}}{F_{\rm cont}}\Bigr) + veil}{1+veil}\,.
\end{equation}

This procedure could be applied only to 5 accretors in our sample. Unfortunately, 
the low resolution of the spectra acquired by \cite{gandolfietal2008} and the very 
low $S/N$ ratio of some FLAMES spectra were not sufficient to apply our method.

\begin{figure*}	
\begin{center}
\hspace{-.4cm}
\includegraphics[width=9.5cm]{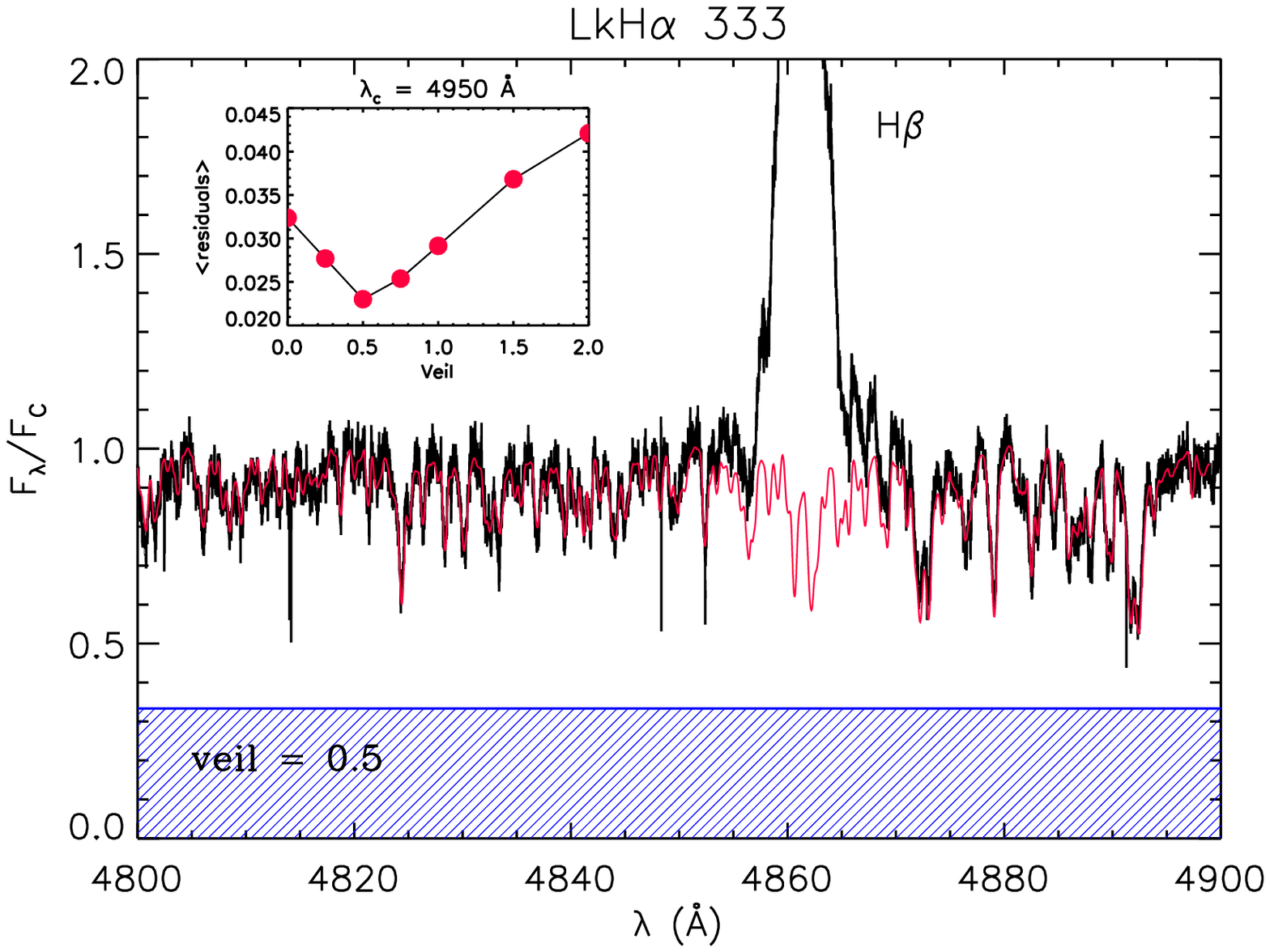}
\hspace{-.6cm}
\includegraphics[width=9.5cm]{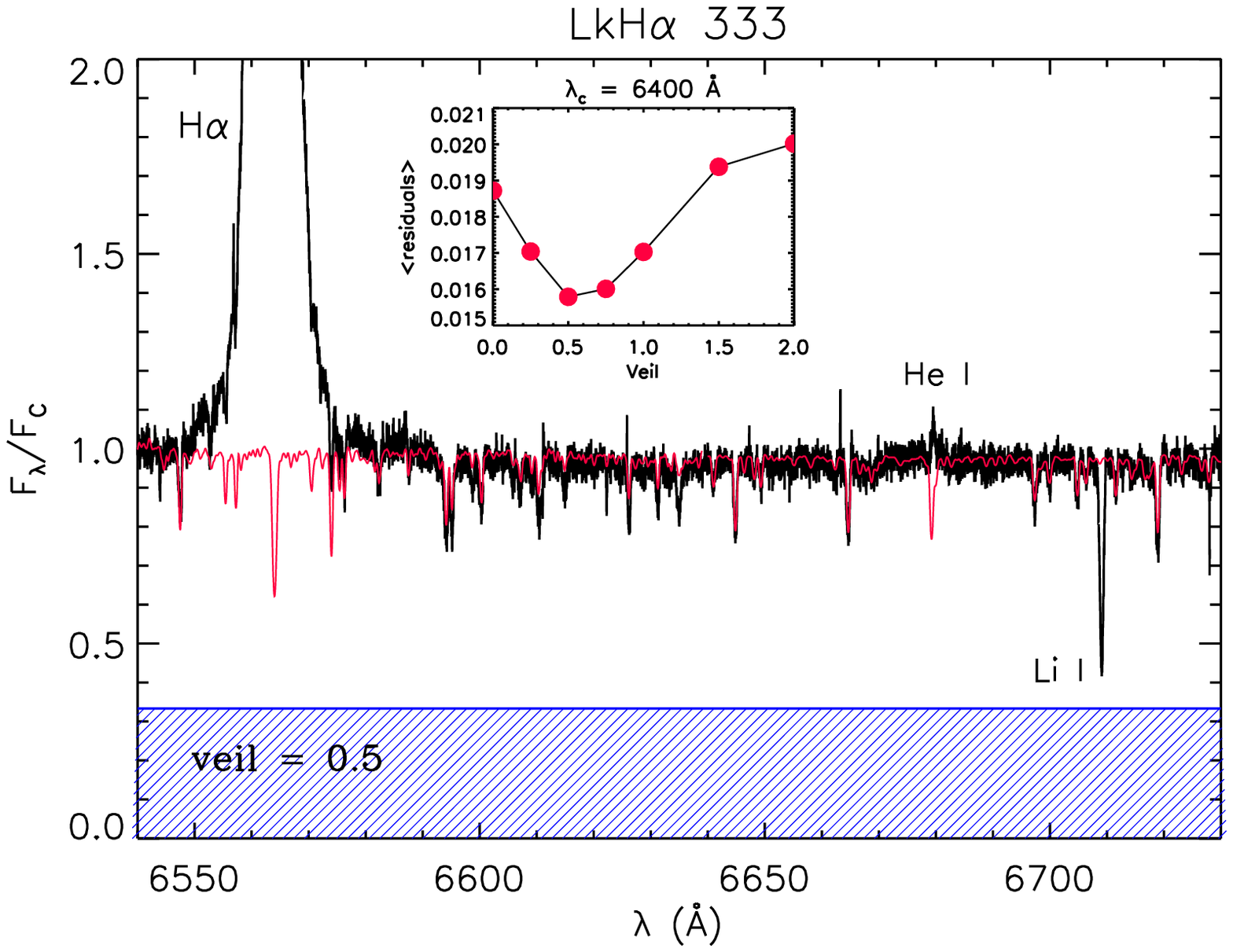}
\vspace{-.3cm}
\caption{Example of UVES spectrum of the accreting star LkH$\alpha$~333 both in the H$\beta$ ({\it left panel}) 
and in the H$\alpha$ ({\it right panel}) region. In both panels, the black solid line represents the spectrum of the target, 
while the red solid line is the spectrum of the best template reproducing the target at $veil=0.5$. 
The residuals as a function of the $veil$ parameter are plotted in the inset. The hatched area is the level 
of the veiling at 0.5. }
\label{fig:lkha_veiling}
\end{center}
\end{figure*}

In Fig.~\ref{fig:lkha_veiling} we show an example of an accreting star observed with 
UVES (LkH$\alpha$~333), with $veil = 0.5$, as found by ROTFIT. In Table~\ref{tab:veiling}, we 
list the mean veiling derived from the FLAMES spectra. Using these values, we could 
estimate the new $\dot M_{\rm acc}$ correcting the measured EWs of the lines by the factor 
$(1 + veil)$. We can conclude that the correction for the veiling leads to a difference of 
$\sim$ 0.25 dex in $\log \dot M_{\rm acc}$ at most, i.e. within the errors of our estimates 
and not affecting our conclusions. Similar results were also found by \cite{costiganetal2012}. 
Hereafter, as we could not evaluate the veiling for all our targets, we will adopt the $\dot M_{\rm acc}$ 
without any correction for the veiling.

\setlength{\tabcolsep}{2.5pt}
\begin{table}[h]
\caption{Mean veiling and difference in $\log \dot M_{\rm acc}$ for the targets for which we 
could run the ROTFIT code.} 
\label{tab:veiling}
\begin{center}
\begin{tabular}{lcc}
\hline
\hline
ID Name    & $veil$ &  $\Delta \log \dot M_{\rm acc}$\\ 
	   &        & (dex) \\ 
\hline
TTS~J050646.1$-$031922& 0.50 & 0.25 \\
RX~J0506.9$-$0319~SE  & 0.25 & 0.10 \\
LkH$\alpha$~333       & 0.50 & 0.20 \\
L1616~MIR~4	      & 0.50 & 0.20 \\
RX~J0507.1$-$0321     & 0.50 & 0.20  \\
\hline
\end{tabular}
\end{center}
\end{table}
\normalsize

\subsection{Variability}
\label{sec:variability}
Being based on single ``epoch'' measurements of line equivalent widths and continuum fluxes estimates, our 
calculations of line luminosity and accretion luminosity represent only an instantaneous snapshot of $L_{\rm acc}$ 
and $\dot M_{\rm acc}$. As in previous investigations in other star forming regions (SFRs; 
see, e.g., \citealt{nguyenetal2009, biazzoetal2012, costiganetal2012, 
fangetal2013, costiganetal2014}), and based on multi-epoch observations of several of our targets, we estimate that 
short-time scale ($\sim$48 hours) variations may induce a scatter on $\log \dot M_{\rm acc}$ of $< 0.3$\,dex, while at a 
longer time scale (a few years) it may be up to $\sim 0.6$\,dex (see Appendix\,\ref{sec:TTS050649.8-032104}). Therefore, as claimed 
in those studies, here we also conclude that YSOs variability may account for variations in $\log \dot M_{\rm acc}$ in 
the range of $\sim 0.2-0.6$\,dex.

\section{Results and discussion}
\label{sec:discussion}

In the following, we discuss the accretion properties of the sample and their link
with the stellar parameters and the IR colors.

\subsection{Accretion luminosity versus stellar parameters}
\label{sec:accretion_luminosity}
\begin{figure}[t!]
\begin{center}
 \begin{tabular}{c}
\hspace{-.6cm}
\includegraphics[width=9.5cm]{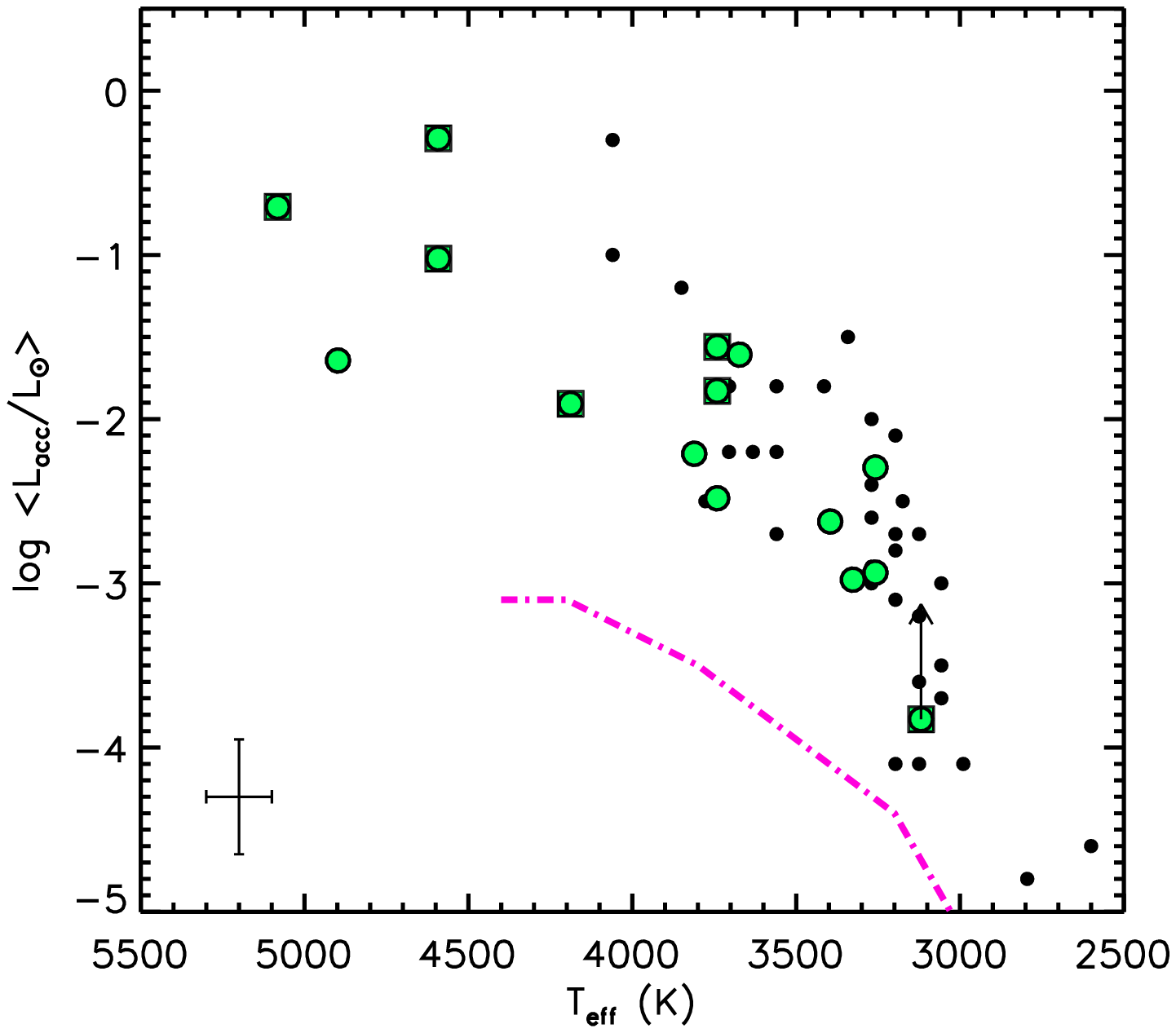}
\vspace{-.4cm}
 \end{tabular}
\caption{Mean accretion luminosity versus effective temperature. 
The dash-dotted line marks the locus below which chromospheric emission is important
in comparison with $L_{\rm acc}$ \citep{manaraetal2013}. The vertical arrow represents 
the position of the sub-luminous object after applying the correction factor 
described in Appendix~\ref{sec:sub-luminous}. Small filled dots represent the \cite{alcalaetal2014} 
sample of low-mass stars in the Lupus SFR. Mean error bars are overplotted on the 
lower-left corner of the panel. Symbols are as in Fig.~\ref{fig:2MASS_color_color}. 
}
\label{fig:Lacc_Teff_mean} 
 \end{center}
\end{figure}

At low levels of accretion, the chromospheric emission may have an important 
impact on the estimates of $L_{\rm acc}$  (see \citealt{manaraetal2013}, and references 
therein). This contribution should be therefore considered when accretion properties are 
studied. As shown in Fig.~\ref{fig:Lacc_Teff_mean}, the accretion luminosity of the YSOs 
in L1615/L1516 decreases monotonically with the effective temperature. The dashed line in 
this figure shows the chromospheric level as determined by \cite{manaraetal2013}, 
and represents the locus below which the contribution of chromospheric emission starts to be 
important in comparison with energy losses due to accretion. All accreting YSOs in L1615/L1616 
fall well above the ``systematic noise'' due to chromospheric emission and show $L_{\rm acc}/L_{\odot}$ 
very similar to the values recently derived for members of the Lupus SFR by 
\cite{alcalaetal2014} and estimated through primary diagnostics. 

\begin{figure}[t!]
\begin{center}
 \begin{tabular}{c}
\hspace{-.6cm}
\includegraphics[width=9.5cm]{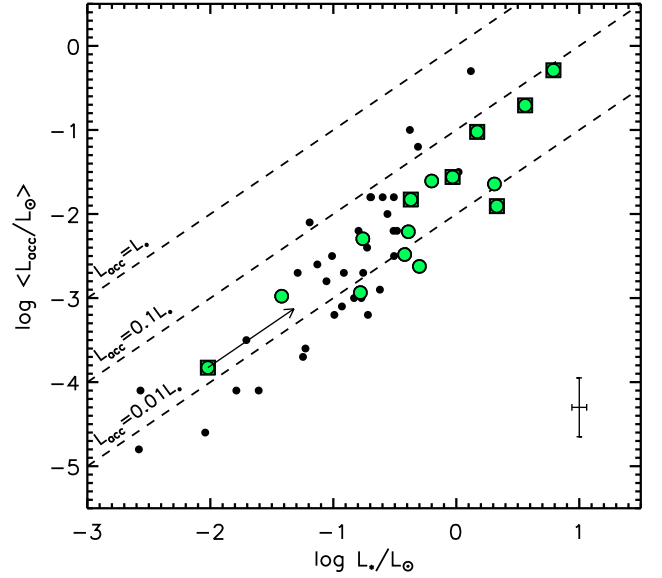}
\vspace{-.4cm}
 \end{tabular}
\caption{Mean accretion luminosity versus stellar luminosity. The dashed lines represent 
the loci of the three $L_{\rm acc}-L_{\star}$ relations, as labeled. The arrow represents 
the position of the sub-luminous object after applying the correction factor 
described in Appendix~\ref{sec:sub-luminous}. The Lupus YSOs by \cite{alcalaetal2014} are 
overlaid as small filled dots. Mean error bars are overplotted on the lower-right corner of 
the panel. Symbols are as in Fig.~\ref{fig:2MASS_color_color}. 
}
\label{fig:Lacc_Lstar_mean} 
 \end{center}
\end{figure}

Figure~\ref{fig:Lacc_Lstar_mean} shows the mean accretion luminosity as a function of 
the stellar luminosity. As already observed by previous investigations in other SFRs, like 
$\rho$~Ophiucus, Taurus, and Lupus (\citealt{muzerolleetal1998, nattaetal2006, alcalaetal2014}), 
the accretion luminosity increases with the stellar luminosity. In our sample of accreting 
stars, $L_{\rm acc}$ follows a trend which is similar to the recent power-law found 
by \cite{alcalaetal2014} in the Lupus star-forming region. Moreover, the dispersion of 
our data points in $L_{\rm acc}$ is similar. As in other star forming regions, the accretion 
luminosity of the YSOs in L1615/L1616 is a fraction of the stellar luminosity, and falls in the 
range between 0.1$L_\star$ to 0.01$L_\star$ (see, e.g., \citealt{muzerolleetal1998, whitehillenbrand2004, 
antoniuccietal2011, carattiogarattietal2012, alcalaetal2014}).

\subsection{Mass accretion rate versus stellar mass}
\label{sec:accretion_parameters}

The distribution of YSOs in the $\dot M_{\rm acc}$ versus $M_\star$ plane provides
an important diagnostic for the studies of the evolution of mass accretion
(see \citealt{hartmannetal2006}). The $\dot M_{\rm acc}$ versus $M_\star$ relationship
has been obtained for a number of different star-forming regions (e.g., Taurus, Ophiuchus, 
$\sigma$~Orionis, Orion Nebula Cluster, Trumpler~37). In all regions studied so far 
it has been found that, although there is a rough correlation of $\dot M_{\rm acc}$ with 
the square of $M_\star$, the scatter of $\dot M_{\rm acc}$ for a given mass is very large 
(e.g., \citealt{muzerolleetal2005, nattaetal2006}). 

The physical origin of the $\dot M_{\rm acc}\propto M_\star^\alpha$ relationship, 
with $\alpha \approx 2$, is still unclear. 
\cite{alexanderarmitage2006} have suggested that the correlation  reflects the 
initial conditions  established  when the disk formed, followed by subsequent 
viscous disk evolution of the disk. The natural decline of the mass accretion rate 
with age in viscous disk evolution and effects due to evolutionary differences within a sample have 
been ruled out as possible cause for the large spread of the relationship within individual 
star forming regions (\citealt{mohantyetal2005, nattaetal2006}). Moreover, 
short-term (see, e.g., \citealt{nguyenetal2009, biazzoetal2012}) and long-term 
variability may contribute to, but cannot explain the large vertical spread of the $\dot M_{\rm acc}-M_\star$ 
relationship (\citealt{biazzoetal2012, costiganetal2012,costiganetal2014}). It appears more 
likely related to a spread in the properties of the parental 
cores, their angular momentum in particular (e.g., \citealt{hartmannetal2006, 
dullemondetal2006}), in stellar properties, such as X-ray emission 
(\citealt{muzerolleetal2003}), or on the competition between different accretion 
mechanisms, such as viscosity and gravitational instabilities 
(\citealt{vorobyovandbasu2008}). As opposed to \cite{dullemondetal2006}, 
\cite{ercolanoetal2014} argued that the $\dot M_{\rm acc}$ versus $M_\star$ relation 
for a population of disks dispersing via X-ray photo-evaporation is completely 
determined by the shape of the X-ray luminosity function, hence requires no 
spread in initial conditions other than the dependence on stellar mass. On the 
other hand, \cite{alcalaetal2014} have concluded that mixing mass-accretion 
rates calculated with different techniques may increase the scatter in the
$\dot M_{\rm acc}$ versus $M_\star$ relationship. They also have claimed that the 
different methodologies used to derive accretion luminosity and line luminosity, 
as well as the different evolutionary models used to estimate masses may lead 
to significantly different results on the slope of the relationship. 

\begin{figure*}[h!]
\begin{center}
 \begin{tabular}{c}
\includegraphics[width=18cm]{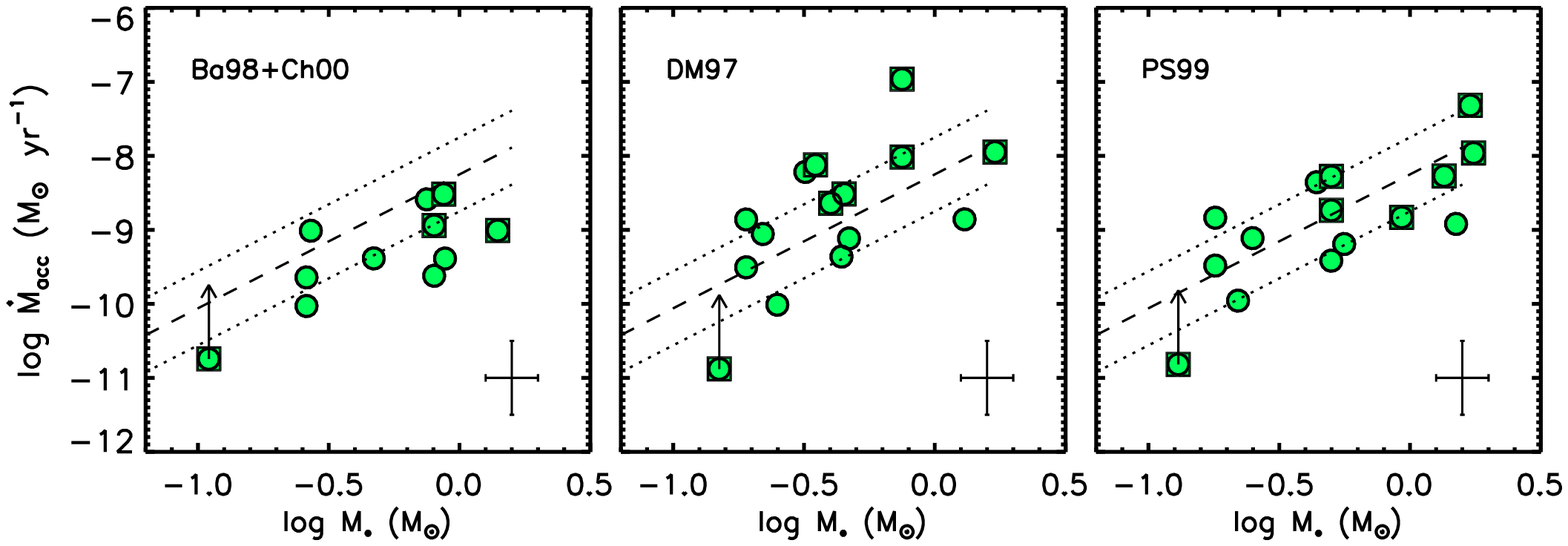}
\vspace{-.2cm}
 \end{tabular}
\caption{Mass accretion rate versus stellar mass drawn from the Ba98+Ch00, DM97, and 
PS99 evolutionary tracks, respectively from the left to the right panel. The dashed 
and dotted lines represent the $\dot M_{\rm acc}-M_\star$ relationship and the 
$1 \sigma$ deviation from the fit obtained with X-Shooter@VLT observations 
by \cite{alcalaetal2014} for YSOs in Lupus. The vertical arrow represents 
the position of the sub-luminous object after applying the correction 
factor described in Appendix~\ref{sec:sub-luminous}. Mean error bars are overplotted 
on the lower-right corner of each panel. Symbols are as in Fig.~\ref{fig:2MASS_color_color}.
}
\label{fig:Macc_Mass_tracks} 
 \end{center}
\end{figure*}

The results in the $\dot M_{\rm acc}-M_\star$ plane for our sample of accretors in L1615/L1616 
are shown in Fig.~\ref{fig:Macc_Mass_tracks}. The three panels in this figure correspond
to the values of $\dot M_{\rm acc}$ calculated from the three estimates of 
$M_\star$ drawn from the three evolutionary models, as labeled in the figure.
Since our number statistics is low, we do not attempt a linear fit to the 
$\dot M_{\rm acc}-M_\star$ relationships, but for comparison, we overplot 
the $\dot M_{\rm acc}-M_\star$ linear fit with a slope of $1.8\pm0.2$ recently 
calculated for Lupus YSOs \citep{alcalaetal2014}, for which the accretion luminosity was 
directly derived by modeling the excess emission from the UV to the near-infrared as the 
continuum emission of a slab of hydrogen. Similar findings were also obtained 
in other T associations, like Taurus or Chamaeleon (see, e.g., 
\citealt{herczeghillenbrand2008, antoniuccietal2011, biazzoetal2012}).

The accretors in L1615/L1616 follow closely the $\dot M_{\rm acc}-M_\star$
relationship seen for the YSOs in Lupus by \cite{alcalaetal2014}, but interestingly the scatter 
changes depending on the evolutionary tracks used to derive the stellar masses. As concluded
in Appendix~\ref{sec:Mass_Macc_tracks}, the different evolutionary tracks
have negligible effects on the computation of $\log \dot M_{\rm acc}$, meaning 
that the scatter in the $\log \dot M_{\rm acc}-\log M_\star$ diagram
is mainly induced by the uncertainty on the mass, which is model-dependent.
The DM97 tracks seem to produce the largest scatter. 

We stress, however, that some scatter in the $\dot M_{\rm acc}-M_\star$ relationship 
(up to around $\pm0.5-0.6$ dex in $\log \dot M_{\rm acc}$; see, e.g., \citealt{costiganetal2014}, and 
references therein) may come from intrinsic variability, as our line luminosity determinations were 
obtained from single ``epoch'' measurements of line equivalent widths and 
assuming continuum flux coming from model atmospheres (see Section\,\ref{sec:variability}).

\subsection{Accretion versus infrared properties}
\label{sec:color_Macc}

Near- and mid-IR colors can be used to probe the inner disk region. 
\cite{hartiganetal1995}, studying a sample of 42 T-Tauri stars and  using 
the $K-L$ color excesses, pointed out that disk dissipation is mainly 
due to the formation of micron-sized dust particles, which merge together to create 
planetesimals and protoplanets at the end of the CTTs phase. Protoplanets may 
clear the innermost part of the disk where the gas and dust have temperature of the 
order of $\sim$ 1000 K and emit mainly at near-IR and mid-IR wavelengths. This causes 
the disk to decrease or loose its near-IR color excess and at the same time the opening of a 
gap in the disk (see, e.g., \citealt{linpapaloizou1993}), thereby possibly terminating 
accretion from the disk onto the star.

\begin{figure*}	
\begin{center}
 \begin{tabular}{c}
\includegraphics[width=7.cm]{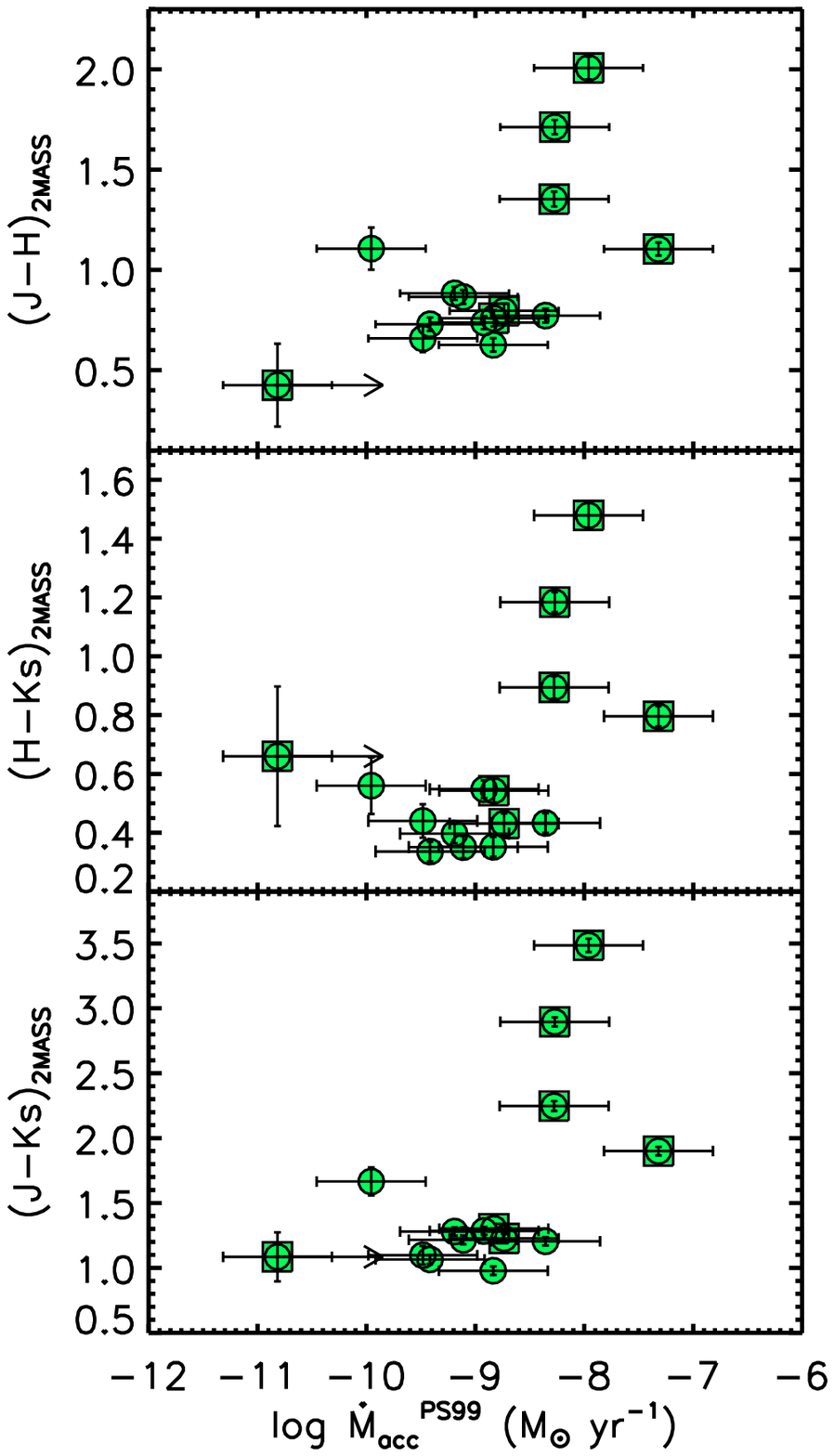}
\hspace{.8cm}
\includegraphics[width=7.cm]{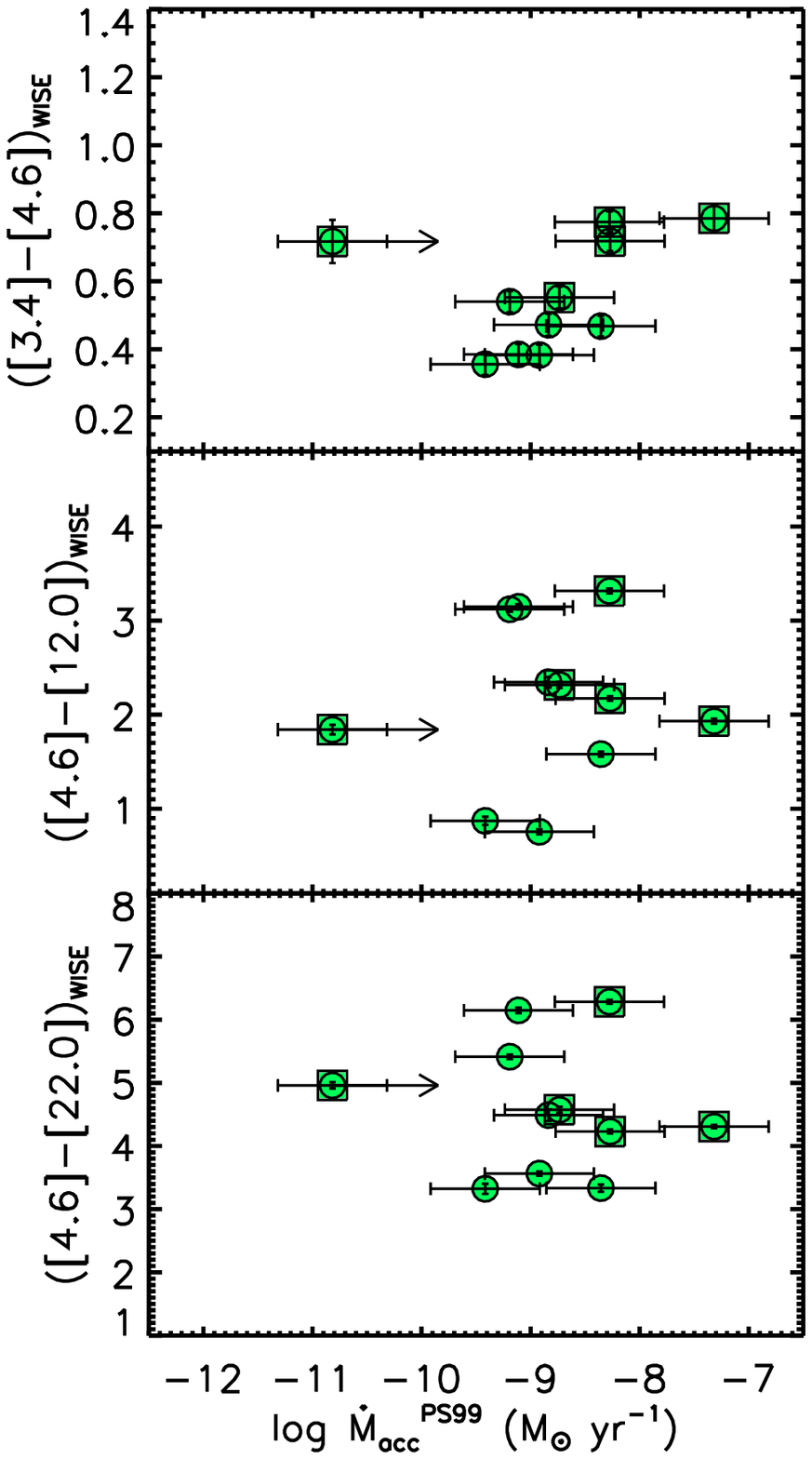}
 \end{tabular}
\caption{{\it 2MASS} ({\it left panel}) and {\it WISE} ({\it right panel}) colors versus mass 
accretion rates derived using the PS99 stellar masses. The horizontal arrows represent the position
of the sub-luminous object after correction of $\dot M_{\rm acc}$, as explained in 
Appendix~\ref{sec:sub-luminous}. Symbols are as in Fig.~\ref{fig:2MASS_color_color}.}
\label{fig:macc_2MASSWISE_colors}
 \end{center}
\end{figure*}

With the aim of investigating possible relationships between IR colors and 
accretion properties, we plotted in Fig.~\ref{fig:macc_2MASSWISE_colors} 
the $J-H$, $H-K{\rm s}$, $J-K{\rm s}$ $2MASS$ colors and the [3.4]$-$[4.6], 
[4.6]$-$[12.0], [4.6]$-$[22.0] $WISE$ colors as a function of the mass 
accretion rates derived using the PS99 tracks, from which masses could 
be estimated for all the accretors we studied in this work. Despite the poor statistics, 
the behavior of the $2MASS$ and $WISE$ colors with accretion is different. 
While the $2MASS$ colors tend to rise at $\dot M_{\rm acc} \gtsim 10^{-10} M_\odot$\,yr$^{-1}$, 
the $WISE$ ones show no trend with $\dot M_{\rm acc}$.
In order to quantify the degree of possible correlations in 
Fig.~\ref{fig:macc_2MASSWISE_colors}, we computed the Spearman's rank correlation 
coefficients with the IDL platform. 
These correlation coefficients, ranging from 0 to 1, show values around 0.6 
for the relations between $\dot M_{\rm acc}$ and $2MASS$ colors, and values 
very close to zero for the $WISE$ colors, meaning that the $2MASS$ colors 
show an increasing trend with $\log \dot M_{\rm acc}$, while no trend is 
detected for the relations between $\dot M_{\rm acc}$ and the $WISE$ colors. 

The trend between $2MASS$ colors and $\log \dot M_{\rm acc}$ could be an 
indication that objects with strong accretion have optically thick inner 
disks, as found in previous works (\citealt{hartiganetal1995, rigliacoetal2011a, biazzoetal2012}). 
In particular, we can identify the regions 
$\log \dot M_{\rm acc} \gtsim -8.5$ dex and $J-H \gtsim$ 1.0 
or $H-K{\rm s} \gtsim$ 0.7 or $J-K{\rm s} \gtsim$ 1.75 as those where strong accreting 
objects with large near-IR excess are found in L1615/L1616.

\section{Conclusions}
\label{sec:conclusions}
In this paper, we investigated the accretion and the IR properties of YSOs
in the cometary cloud L1615/L1616 in Orion. For this purpose we used
intermediate resolution (FLAMES@VLT) and low-resolution (FORS2@VLT + VIMOS@VLT) 
optical spectroscopy for 23 and 31 objects, respectively. Our main results can be 
summarized as follows:

\begin{enumerate}
\item The YSOs in L1615/L1616 observed with FLAMES show a narrow distribution in 
radial velocity peaked at $23.2\pm3.1$ km s$^{-1}$, showing they are dynamically associated with 
the cloud, and mean lithium abundance of $3.3\pm0.3$\,dex 
confirming their membership to the cometary cloud and their youth.
\item The fraction of accretors in L1615/L1516 is close to 30\%, consistent with 
the fraction of disks recently reported by \cite{ribasetal2014} for an average age of 3\,Myr.
\item The mass accretion rates ($\dot M_{\rm acc}$) derived through several secondary 
diagnostics (H$\alpha$, H$\beta$, \ion{He}{i} $\lambda$5876\,\AA, \ion{He}{i} $\lambda$6678\,\AA, 
and \ion{He}{i} $\lambda$7065\,\AA) are in the range $\sim 10^{-10}-10^{-7} M_\odot$\,yr$^{-1}$ 
for stars with $\sim 0.1-2.3 M_\odot$. These accretion rates are similar 
to those of YSOs of similar mass in other star forming regions.
\item The accretion properties of the YSOs in L1615/L1616 have the same behavior 
as YSOs in other star-forming regions, like Lupus or Taurus. This might imply that environmental 
conditions at which the cometary cloud is exposed uninfluenced the accretion evolution of the YSOs 
in this cometary cloud.
\item As recently found by other authors, we confirm that different methods used 
to derive stellar parameters and mass accretion rates introduce dispersion 
in the $\dot M_{\rm acc} - M_\star$ relation; in particular, the differences in the 
evolutionary tracks used to derive $M_\star$ and then $\dot M_{\rm acc}$ produce a scatter 
in the $\dot M_{\rm acc} - M_\star$ relationship, but no significant systematic effect on 
$\dot M_{\rm acc}$.
\item The color$-\dot M_{\rm acc}$ diagrams suggest that strong accretors (i.e. those with 
$\log \dot M_{\rm acc} \gtsim -8.5$ dex) show large excesses in the $JHK{\rm s}$ bands, indicative 
of inner optically thick disk, as in previous studies.
\end{enumerate}
 
\begin{acknowledgements}
The authors are very grateful to the referee for carefully reading the paper and for his/her useful 
remarks that allowed us to improve the previous version of the manuscript. KB acknowledges 
the {\it Osservatorio Astronomico di Capodimonte} for the support given during her visits.
This research made use of the SIMBAD database, operated at the CDS (Strasbourg,
France). This publication makes use of data products 
from the Two Micron All Sky Survey, which is a joint project of the University 
of Massachusetts and the Infrared Processing and Analysis Center/California 
Institute of Technology, funded by NASA and the National Science Foundation. 
This publication makes use of data products from the Wide-field Infrared Survey 
Explorer, which is a joint project of the University of California, Los Angeles, 
and the Jet Propulsion Laboratory/California Institute of Technology, funded by 
the National Aeronautics and Space Administration.
This work was financially supported by the PRIN INAF 2013 "Disks, jets and the dawn of planets".
\end{acknowledgements}

\bibliographystyle{../../aa-package_Apr2013/aa}

\begin{thebibliography}{}
\bibitem[\protect\citeauthoryear{Alcal\'a et al.}{2008}]{alcalaetal2008} Alcal\'a, J. M., Covino, E., \& Leccia, S. 2008, in {\it Handbook of Star Forming Regions}, 
Vol. I: The Northern Sky ASP Monograph Publications, Vol. 4, ed. B. Reipurth, 801 
\bibitem[\protect\citeauthoryear{Alcal\'a et al.}{2014}]{alcalaetal2014} Alcal\'a, J. M., Natta, A., Manara, C., et al. 2014, \aap, 561, A2
\bibitem[\protect\citeauthoryear{Alcal\'a et al.}{2011}]{alcalaetal2011} Alcal\'a, J. M., Stelzer, B., Covino, E., et al. 2011, Astron. Nachr., 332, 242
\bibitem[\protect\citeauthoryear{Alcal\'a et al.}{2004}]{alcalaetal2004} Alcal\'a, J. M., Wachter, S., Covino, E., et al. 2004, \aap, 416, 677
\bibitem[Alexander \& Armitage(2006)]{alexanderarmitage2006} Alexander, R. D., \& Armitage P. J. 2006, \apj, 639, L83 
\bibitem[Antoniucci et al.(2011)]{antoniuccietal2011} Antoniucci, S., Garc\'{i}a-L\'opez, R., Nisini, B., et al. 2011, \aap, 534, 32
\bibitem[Baraffe \& Chabrier(2010)]{baraffechabrier2010} Baraffe, I., \& Chabrier, G. 2010, \aap, 521, 44
\bibitem[Baraffe et al.(1998)]{baraffeetal1998} Baraffe, I., Chabrier, G., Allard, F., \& Hauschildt, P. H. 1998, \aap, 337, 403
\bibitem[Bessell \& Brett(1988)]{bessellbrett1988} Bessell, M. S. \& Brett, J. M. 1988, \pasp, 100, 1134
\bibitem[Biazzo et al.(2009)]{biazzoetal2009} Biazzo, K., Melo, C. H. F., Pasquini, L., et al. 2009, \aap, 508, 1301
\bibitem[Biazzo et al.(2012)]{biazzoetal2012} Biazzo, K., Alcal\'a, J. M., Covino, E., et al. 2012, \aap, 547, A104
\bibitem[Blecha et al.(2000)]{blechaetal2000} Blecha, A., Cayatte, V., North, P., Royer, F., \& Simond, G. 2000, Proc. SPIE, 4008, 467
\bibitem[Brice\~no(2008)]{briceno2008} Brice\~no, C. 2008, in {\it Handbook of Star Forming Regions}, Vol. I: The Northern Sky ASP Monograph 
Publications, Vol. 4, ed. B. Reipurth, 838
\bibitem[Caratti o Garatti et al.(2012)]{carattiogarattietal2012} Caratti o Garatti, A., Garc\'{i}a-L\'opez, R., Antoniucci, S., et al. 2012, \aap, 538, A64
\bibitem[Chabrier et al.(2000)]{chabrieretal2000} Chabrier, G., Baraffe, I., Allard, F., \& Hauschildt, P. H. 2000, \aap, 542, 464
\bibitem[Comer\'on et al.(2003)]{comeronetal2003} Comer\'on, F., Fern\'andez, M., Baraffe, I., Neuh\"auser, R., \& Kaas, A. A. 2003, \aap, 406, 1001
\bibitem[Costigan et al.(2012)]{costiganetal2012} Costigan, G., Sch\"olz, A., Stelzer, B., et al.  2012, \mnras, 427, 1344
\bibitem[Costigan et al.(2014)]{costiganetal2014} Costigan, G., Vink, Jorick S., Sch\"olz, A., Ray, T. \& Testi, L. 2014, \mnras, 440, 3444
\bibitem[Cutri et al.(2003)]{cutrietal2003} Cutri, R. M., Skrutskie, M. F., van Dyk, S., et al. 2003, Explanatory Supplement to the 2MASS All Sky Data Release
\bibitem[Cutri et al.(2012)]{cutrietal2012} Cutri, R. M., Wright, E. L., Conrow, T., et al. 2012, Explanatory Supplement to the WISE All-Sky Data Release Products
\bibitem[D'Antona \& Mazzitelli(1997)]{dantonamazzitelli1997} D'Antona, F., \& Mazzitelli, I. 1997, MSAIt, 68, 807
\bibitem[Dullemond et al.(2006)]{dullemondetal2006} Dullemond, C. P., Natta, A., \& Testi, L. 2006, \apj, 645, 69
\bibitem[Ercolano et al.(2014)]{ercolanoetal2014} Ercolano, B., Mayr, D., Owen, J. E., Rosotti, G., \& Manara, C. F. 2014, \mnras, 439, 256
\bibitem[Fang et al.(2009)]{fangetal2009} Fang, M., van Boekel, R., Wang, W., et al. 2009, \aap, 504, 461
\bibitem[Fang et al.(2013)]{fangetal2013} Fang, M., Kim, J. S., van Boekel, R., et al. 2013, \apjs, 207, 5
\bibitem[Frasca et al.(2003)]{frascaetal2003} Frasca, A., Alcal\'a, J. M., Covino, E., et al. 2003, \aap, 405, 149
\bibitem[Frasca et al.(2014)]{frascaetal2014} Frasca, A., Biazzo, K., Lanzafame, A., et al. 2014, \aap, submitted
\bibitem[Frasca et al.(2006)]{frascaetal2006} Frasca, A., Guillout, P., Marilli, E., et al. 2006, \aap, 454, 301
\bibitem[Gandolfi et al.(2008)]{gandolfietal2008} Gandolfi, D., Alcal\'a, J. M., Leccia, S., et al. 2008, \apj, 687, 1303
\bibitem[Gullbring et al.(1998)]{gullbringetal1998} Gullbring, E., Hartmann, L., Brice\~no, C., \& Calvet, N. 1998, \apj, 492, 323
\bibitem[Hartigan et al.(1995)]{hartiganetal1995} Hartigan, P., Edwards, S., \& Ghandour, L. 1995, \apj, 452, 736
\bibitem[Hartmann(1998)]{hartmann1998} Hartmann, L. 1998: in {\it Accretion Processes in Star Formation}, Cambridge Univ. Press
\bibitem[Hartmann et al.(2006)]{hartmannetal2006} Hartmann, L., D'Alessio, P., Calvet, N., \& Muzerolle, J. 2006, \apj, 648, 484
\bibitem[Hauschildt et al.(1999)]{hauschildtetal1999} Hauschildt, P. H., Allard, F., \& Baron, E. 1999, \apj, 512, 377
\bibitem[Herczeg \& Hillenbrand(2008)]{herczeghillenbrand2008} Herczeg, G. J., \& Hillenbrand, L. A. 2008, \apj, 681, 594
\bibitem[Ingleby et al.(2013)]{ingleby2013} Ingleby, L., Calvet, N., Herczeg, G., et al. 2013, \apj, 767,112
\bibitem[Kenyon \& Hartmann(1995)]{kenyonhartmann1995} Kenyon, S. J., \& Hartmann, L. 1995, \apjs, 101, 117
\bibitem[Koenig et al.(2012)]{koenigetal2012} Koenig, X. P., Leisawitz, D. T., Benford, D. J., et al. 2012, \apj, 744, 130
\bibitem[K\"onigl (1991)]{konigl91} K\"onigl, A. 1991, \apj, 370, L39
\bibitem[Lee \& Chen(2007)]{leechen2007} Lee, H.-T., \& Chen, W. P., 2007, \apj, 657, 884
\bibitem[Lin \& Papaloizou(1993)]{linpapaloizou1993} Lin, D. N. C., \& Papaloizou, J. C. B. 1993: in {\it Protostars and planets III}, 
University of Arizona Press, E. H. Levy, \& J. I. Lunine eds., p. 749
\bibitem[Looper et al.(2010)]{looperetal10} Looper, D., Mohanty, S., Bochanski, J. J., et al. 2010, \apj, 714, 46
\bibitem[Lynds(1962)]{lynds1962} Lynds, B. T. 1962, \apjs, 7, 1
\bibitem[Manara et al.(2013)]{manaraetal2013} Manara, C. F., Testi, L., Rigliaco, E., et al. 2013, \aap, 551, 107
\bibitem[Meyer et al.(1997)]{meyeretal1997} Meyer, M. R., Calvet, N., \& Hillenbrand, L. A. 1997, \aj, 114, 288
\bibitem[Modigliani et al.(2004)]{modiglianietal2004} Modigliani, A., Mulas, G., Porceddu, I., et al. 2004, The Messenger, 118, 8
\bibitem[Mohanty et al.(2005)]{mohantyetal2005} Mohanty, S., Jayawardhana, R., \& Basri, G., 2005, \apj,626, 498
\bibitem[Muzerolle et al.(1998)]{muzerolleetal1998} Muzerolle, J., Hartmann, L., \& Calvet, N. 1998b, \aj, 116, 2965
\bibitem[Muzerolle et al.(2003)]{muzerolleetal2003} Muzerolle, J.,  Hillenbrand, L., Calvet, N., Brice\~no, C., Hartmann, L. 2003, \apj, 592, 266
\bibitem[Muzerolle et al.(2005)]{muzerolleetal2005} Muzerolle, J., Luhman, K., Brice\~no, C., et al. 2005, \apj, 625, 906
\bibitem[Natta et al.(2006)]{nattaetal2006} Natta, A., Testi, L., \& Randich, S. 2006, \aap, 452, 245
\bibitem[Nguyen et al.(2009)]{nguyenetal2009} Nguyen, D. C., Scholz, A., van Kerkwijk, M. H., Jayawardhana, R., Brandeker, A., 2009, 
ApJ, 694, L153
\bibitem[Palla et al.(2007)]{pallaetal2007} Palla, F., Randich, S., Pavlenko, Y. V., Flaccomio, E., \& Pallavicini, R. 2007, \apj, 659, 41
\bibitem[Palla \& Stahler(1999)]{pallastahler1999} Palla, F., Stahler, S. W. 1999, \apj, 525, 772
\bibitem[Pavlenko \& Magazz\`u(1996)]{pavlenkomagazzu1996} Pavlenko, Y. V., \& Magazz\`u, A. 1996, \aap, 311, 961
\bibitem[Ribas et al.(2014)]{ribasetal2014} Ribas, A., Mer\'{i}n, B., Bouy, H., Maud, L. T. 2014, \aa, 561, 54
\bibitem[Rigliaco et al.(2011a)]{rigliacoetal2011a} Rigliaco, E., Natta, A., Randich, S., Testi, L., \& Biazzo, K. 2011a, \aap, 525, 47
\bibitem[Rigliaco et al.(2011b)]{rigliacoetal2011b} Rigliaco, E., Natta, A., Randich, S., et al. 2011b, \aap, 526, 6
\bibitem[Rigliaco et al.(2012)]{rigliacoetal2012} Rigliaco, E., Natta, A., Testi, L., et al. 2012, \aap, 548, 56
\bibitem[Schisano et al.(2009)]{schisanoetal2009} Schisano, E., Covino, E., Alcal\'a, et al. 2009, \aap, 501, 3
\bibitem[Sergison et al.(2013)]{sergisonetal2013} Sergison, D. J., Mayne, N. J., Naylor, R., Jeffries, R. D., \& Bell, P. M. 2013, \mnras, 434, 966
\bibitem[Shu et al.(1994)]{shu94} Shu, F., Najita, J., Ostriker, E., \& Wilkin, F. 1994, \apj, 429, 781
\bibitem[Stanke et al.(2002)]{stankeetal2002} Stanke, R., Smith, M. D., Gredel, R., \& Szokoly, G. 2002, \apj, 393, 251
\bibitem[Tonry \& Davis(1979)]{tonrydavis1979} Tonry, J., \& Davis, M. 1979, \apj, 84, 1511
\bibitem[Uchida \& Shibata(1985)]{uchida85} Uchida, Y. \& Shibata, K. 1985, PASJ, 37, 515
\bibitem[Vorobyov \& Basu(2008)]{vorobyovandbasu2008} Vorovyov, E. J., \& Basu, S. 2008, \apj, 676, 139
\bibitem[White \& Basri(2003)]{whitebasri2003} White, R. J., Basri, G. 2003, \apj, 582, 1109
\bibitem[White \& Hillenbrand(2004)]{whitehillenbrand2004} White, R. J., \& Hillenbrand, L. A. 2004, \apj, 616, 998
\end{thebibliography}

\Online
\scriptsize
\begin{longtable}{lccccccc}
\caption[ ]{\label{tab:all_param} Observing log, radial velocity, and lithium content for the stars observed with FLAMES. }\\
\hline\hline
ID name & $JD$ & Instrument & \# & $V_{\rm rad}$& $EW_{\rm Li}$ &$\log n({\rm Li})$& Comment\\
	  & ($+2\,450\,000$) & & obs. & (km~s$^{-1}$)  & (m\AA)  & (dex) & \\
\hline
TTS~J050646.1$-$031922&3797.0722 & GIRAFFE &2&    26.2$\pm$6.6  & 587& 3.70  &N1,N2;S1,S2\\
		      &3797.1097 &    "    & &    25.9$\pm$6.9  & 594&       &N1,N2;S1,S2\\   
RX~J0506.8$-$0318     &3795.0535 & GIRAFFE &3&    21.6$\pm$0.7  & 604& 3.15  &  \\
		      &3795.1009 &    "    & &    21.7$\pm$2.7  & 605&       &  \\
		      &3796.0344 &    "    & &    22.6$\pm$0.3  & 581&       &  \\
TTS~J050647.5$-$031910&3795.0535 & GIRAFFE &3&      ... 	& ...& ...   &LSN\\
		      &3795.1009 &    "    & &    28.6$\pm$5.3  & ...&       &   \\
		      &3796.0344 &    "    & &      ... 	& ...&       &LSN\\
RX~J0506.8$-$0327     &3795.0535 & GIRAFFE &2&    26.3$\pm$2.3  & 566& 3.50  &   \\
		      &3795.1009 &    "    & &    26.8$\pm$1.8  & 550&       &   \\
RX~J0506.8$-$0305     &3796.0344 & GIRAFFE &1&    19.1$\pm$4.0  & 565& 3.35  &   \\
TTS~J050649.8$-$032104&3795.1009 & GIRAFFE &2&      ... 	& ...&       &LSN\\
		      &3796.0344 &    "    & &      ... 	& ...&       &LSN\\
RX~J0506.9$-$0319~NW  &3795.1009 & GIRAFFE &1&    21.3$\pm$2.2  & 412& 2.70  &   \\
RX~J0506.9$-$0319~SE  &3797.0722 & GIRAFFE &2&    27.4$\pm$5.8  & 565& 3.40  &N1,N2;S1,S2\\
		      &3797.1097 &    "    & &    27.2$\pm$4.7  & 601&       &N1,N2;S1,S2\\
RX~J0506.9$-$0320~W   &3795.0535 & GIRAFFE &1&    21.4$\pm$1.6  & 562& 3.35  &\\
RX~J0506.9$-$0320~E   &3796.0344 & GIRAFFE &4&    22.9$\pm$3.3  & 409& 3.70  &\\
		      &3797.0298 &    "    & &    23.5$\pm$3.3  & 444&       &     \\
		      &3797.0722 &    "    & &    24.3$\pm$4.4  & 385&       &     \\
		      &3797.1096 &    "    & &    23.9$\pm$5.6  & 377&       &     \\
LkH$\alpha$~333       &3796.0344 & GIRAFFE &1&    25.9$\pm$1.6  & 447& 3.50  &     \\
		      &3795.0535 & UVES    &6&    27.0$\pm$0.5  & 508&       &O1,O2; T\\
		      &3795.1009 &    "    & &      ... 	& ...&       &LSN\\
		      &3796.0767 &    "    & &    27.8$\pm$0.5  & 462&       &O1,O2 \\
		      &3797.0298 &    "    & &    28.4$\pm$0.6  & 455&       &O1,O2 \\
		      &3797.1097 &    "    & &    28.2$\pm$2.0  & 437&       &O1,O2 \\
		      &3797.0723 &    "    & &    27.9$\pm$1.1  & 455&       &O1,O2 \\
L1616~MIR~4	      &3795.1009 & GIRAFFE &4&    24.3$^{\it a}$ & 365& 3.50  &\\
		      &3797.0298 &    "    & &    29.4$\pm$7.9  & ...&       &LSN \\
		      &3797.0722 &    "    & &    28.1$^{\it a}$ & ...&       &LSN  \\
		      &3797.1097 &    "    & &      ... 	& ...&       &LSN  \\
RX~J0507.0$-$0318     &3795.0535 & GIRAFFE &3&    23.1$\pm$2.7  & 556&$>$3.50& \\
		      &3795.1009 &    "    & &    23.7$\pm$1.9  & 583&       & T \\
		      &3796.0344 &    "    & &    23.6$\pm$1.7  & 601&       & \\
RX~J0507.1$-$0321     &3795.0535 & GIRAFFE &2&    20.9$\pm$1.1  & 575&$>$3.50& \\
		      &3796.0344 &    "    & &    21.9$\pm$1.8  & 550&       & \\
RX~J0507.2$-$0323     &3796.0767 & GIRAFFE &1&    24.8$\pm$1.7  & 525& 3.40  & T \\
RX~J0507.3$-$0326     &3795.0535 & GIRAFFE &2&    21.4$\pm$1.1  & 549& 3.40  & \\
		      &3796.0767 &    "    & &    21.8$\pm$1.8  & 545&       & \\
TTS~J050717.9$-$032433&3795.0535 & GIRAFFE &2&    20.2$^{\it a}$ & 524& 3.50  & \\
		      &3796.0767 &    "    & &    20.4$^{\it a}$ & 530&       & \\
RX~J0507.4$-$0320     &3795.0535 & GIRAFFE &3&    20.3$^{\it a}$ & 587& 3.50  & \\
		      &3796.0767 &    "    & &    19.8$^{\it a}$ & 518&       & \\
		      &3796.0344 &    "    & &    19.9$^{\it a}$ & 608&       & \\
RX~J0507.4$-$0317     &3795.0535 & GIRAFFE &3&    19.8$\pm$6.4  & ...& 3.20  &LSN\\
		      &3796.0767 &    "    & &    19.2$^{\it a}$ & 457&       & \\
		      &3796.0344 &    "    & &    20.0$^{\it a}$ & 503&       & \\
TTS~J050729.8$-$031705&3795.0535 & GIRAFFE &3&      ... 	& ...& 3.00  &LSN \\
		      &3796.0767 &    "    & &      ... 	& ...&       &LSN \\
		      &3796.0344 &    "    & &      ... 	& 580&       & \\
TTS~J050730.9$-$031846&3795.0535 & GIRAFFE &2&      ... 	& ...& ...   &LSN \\
		      &3796.0344 &    "    & &      ... 	& ...&       &LSN \\
TTS~J050734.8$-$031521&3795.0535 & GIRAFFE &3&      ... 	& ...& ...   &LSN \\
		      &3796.0767 &    "    & &      ... 	& ...&       &LSN \\
		      &3796.0344 &    "    & &      ... 	& ...&       &LSN \\
RX~J0507.6$-$0318     &3795.0535 & GIRAFFE &3&    20.3$\pm$2.7  & 567& 2.75  & \\
		      &3796.0767 &    "    & &    19.2$\pm$1.7  & 524&       & \\
		      &3796.0344 &    "    & &    19.5$\pm$4.1  & 544&       & \\
\hline							  
\end{longtable}
\footnotesize{Notes:
$^{\it a}$ Due to low $S/N$ ratio, few lines, late spectral type, and/or short wavelength coverage, the radial velocity error may be up to 60\%.
LSN: low $S/N$ spectrum; T: template spectrum used for the CCF analysis.
Some optical forbidden lines are observed in emission: 
O1=[\ion{O}{i}] $\lambda$6300.8\,\AA, O2=[\ion{O}{i}] $\lambda$6363.8\,\AA, S1=[\ion{S}{ii}] $\lambda$6715.8\,\AA, 
S2=[\ion{S}{ii}] $\lambda$6729.8\,\AA, N1=[\ion{N}{ii}] $\lambda$6548.4\,\AA, N2=[\ion{N}{ii}] $\lambda$6583.4\,\AA.
}

\topmargin 3 cm
\scriptsize
\begin{longtable}{lccc|cccc}
\caption[ ]{\label{tab:2mass_wise} Near-IR and Mid-IR photometric data.}\\
\hline\hline
   &  &   {\it 2MASS}& &   & {\it WISE} & & \\
ID name  &  $J$  &  $H$  &  $K$s  & [3.4]  & [4.6] & [12.0] & [22.0] \\
	 & (mag) & (mag) & (mag) & (mag) & (mag) & (mag) & (mag) \\
\hline
1RXS~J045912.4$-$033711 & 10.070$\pm$0.022 &  9.616$\pm$0.024 &  9.474$\pm$0.020 &  9.386$\pm$0.023 &  9.395$\pm$0.018 &  9.299$\pm$0.033 & 8.691$^b$\\
1RXS~J050416.9$-$021426 & 10.661$\pm$0.024 & 10.105$\pm$0.024 &  9.984$\pm$0.025 &  9.910$\pm$0.023 &  9.936$\pm$0.020 &  9.812$\pm$0.048 & 8.943$^b$\\
 TTS~J050513.5$-$034248 & 15.120$\pm$0.041 & 14.598$\pm$0.053 & 14.198$\pm$0.065 & 14.011$\pm$0.030 & 13.837$\pm$0.044 &       12.041$^b$ & 8.769$^b$\\
 TTS~J050538.9$-$032626 & 13.287$\pm$0.024 & 12.601$\pm$0.022 & 12.387$\pm$0.024 & 12.196$\pm$0.024 & 12.093$\pm$0.023 & 11.186$\pm$0.159 & 7.745$\pm$0.152\\
      RX~J0506.6$-$0337 & 10.657$\pm$0.022 & 10.304$\pm$0.027 & 10.205$\pm$0.023 & 10.130$\pm$0.024 & 10.138$\pm$0.021 & 10.226$\pm$0.070 & 9.067$^b$\\
 TTS~J050644.4$-$032913 & 13.641$\pm$0.028 & 13.038$\pm$0.031 & 12.779$\pm$0.032 & 12.613$\pm$0.025 & 12.421$\pm$0.025 & 11.411$\pm$0.146 & 8.731$\pm$0.336\\
TTS~J050646.1$-$031922  & 13.218$\pm$0.025 & 11.506$\pm$0.025 & 10.322$\pm$0.023 &  8.866$\pm$0.023 &  8.148$\pm$0.018 &  5.976$\pm$0.014 & 3.918$\pm$0.026\\
RX~J0506.8$-$0318       & 11.849$\pm$0.023 & 11.108$\pm$0.025 & 10.955$\pm$0.023 & 10.835$\pm$0.023 & 10.819$\pm$0.020 &       10.422$^b$ & 7.696$^b$\\
TTS~J050647.5$-$031910  & 14.286$\pm$0.030 & 13.410$\pm$0.030 & 12.999$\pm$0.023 & 12.586$\pm$0.024 & 12.106$\pm$0.023 &  9.229$\pm$0.180 & 5.786$\pm$0.091\\
RX~J0506.8$-$0327       & 11.599$\pm$0.023 & 10.855$\pm$0.027 & 10.595$\pm$0.026 & 10.584$\pm$0.024 & 10.430$\pm$0.020 &  9.877$\pm$0.047 & 8.337$\pm$0.263\\
RX~J0506.8$-$0305       & 12.939$\pm$0.021 & 12.281$\pm$0.025 & 12.058$\pm$0.028 & 11.951$\pm$0.025 & 11.762$\pm$0.023 & 11.868$\pm$0.305 & 9.073$^b$\\
TTS~J050649.8$-$031933  & 12.391$\pm$0.022 & 11.526$\pm$0.027 & 11.175$\pm$0.026 & 10.747$\pm$0.023 & 10.362$\pm$0.020 &  7.215$\pm$0.020 & 4.214$\pm$0.037\\
TTS~J050649.8$-$032104  & 12.699$\pm$0.026 & 11.346$\pm$0.026 & 10.452$\pm$0.028 &  9.845$\pm$0.023 &  9.071$\pm$0.021 &  5.756$\pm$0.016 & 2.787$\pm$0.022\\
TTS~J050650.5$-$032014  & 13.663$\pm$0.026 & 12.971$\pm$0.027 & 12.541$\pm$0.030 &	  ...	    &	     ...       &	...	  &	   ...     \\
TTS~J050650.7$-$032008  & 13.642$\pm$0.057 & 12.984$\pm$0.037 & 12.544$\pm$0.043 &	  ...	    &	     ...       &	...	  &	   ...     \\
RX~J0506.9$-$0319~NW    &      11.324$^a$  & 11.939$\pm$0.072 &      10.007$^a$  &	  ...	    &	     ...       &	...	  &	   ...     \\
RX~J0506.9$-$0319~SE    &      10.809$^a$  & 10.050$\pm$0.041 &       9.507$^a$  &	  ...	    &	     ...       &	...	  &	   ...     \\
RX~J0506.9$-$0320~W     & 10.042$\pm$0.040 &  8.889$\pm$0.042 &  8.393$\pm$0.039 &	  ...	    &	     ...       &	...	  &	   ...     \\
RX~J0506.9$-$0320~E     & 10.366$\pm$0.023 &  8.965$\pm$0.027 &  8.099$\pm$0.026 &	  ...	    &	     ...       &	...	  &	   ...     \\
TTS~J050654.5$-$032046  & 15.984$\pm$0.082 & 14.878$\pm$0.066 & 14.318$\pm$0.070 &	  ...	    &	     ...       &	...	  &	   ...     \\
LkH$\alpha$~333         & 10.343$\pm$0.022 &  9.239$\pm$0.024 &  8.443$\pm$0.025 &  6.981$\pm$0.030 &  6.196$\pm$0.021 &  4.265$\pm$0.013 & 1.888$\pm$0.018\\
L1616~MIR~4	        & 13.010$\pm$0.045 & 11.003$\pm$0.035 &  9.524$\pm$0.025 &	  ...	    &	     ...       &	...	  &	   ...     \\
RX~J0507.0$-$0318       & 11.190$\pm$0.022 & 10.344$\pm$0.027 & 10.036$\pm$0.023 &  9.975$\pm$0.024 &  9.937$\pm$0.020 &  9.103$\pm$0.051 & 6.687$\pm$0.099\\
TTS~J050657.0$-$031640  & 13.040$\pm$0.023 & 12.414$\pm$0.025 & 12.062$\pm$0.023 & 11.781$\pm$0.024 & 11.309$\pm$0.022 &  8.964$\pm$0.052 & 6.820$\pm$0.092\\
TTS~J050704.7$-$030241  & 14.809$\pm$0.036 & 14.208$\pm$0.049 & 13.964$\pm$0.059 & 13.795$\pm$0.029 & 13.569$\pm$0.037 &       12.652$^b$ & 9.104$^b$\\
TTS~J050705.3$-$030006  & 12.517$\pm$0.023 & 11.789$\pm$0.027 & 11.614$\pm$0.025 & 11.594$\pm$0.023 & 11.539$\pm$0.023 &       11.919$^b$ & 8.931$^b$\\
RX~J0507.1$-$0321       & 12.153$\pm$0.025 & 11.357$\pm$0.027 & 10.926$\pm$0.022 & 10.656$\pm$0.023 & 10.104$\pm$0.022 &  7.787$\pm$0.021 & 5.530$\pm$0.043\\
TTS~J050706.2$-$031703  & 14.660$\pm$0.033 & 14.067$\pm$0.041 & 13.820$\pm$0.048 & 13.498$\pm$0.027 & 13.163$\pm$0.032 & 10.891$\pm$0.145 & 8.390$\pm$0.387\\
RX~J0507.2$-$0323       & 11.659$\pm$0.020 & 11.053$\pm$0.025 & 10.911$\pm$0.025 & 10.806$\pm$0.022 & 10.799$\pm$0.021 & 10.012$\pm$0.060 & 7.505$\pm$0.160\\
TTS~J050713.5$-$031722  & 13.918$\pm$0.031 & 12.855$\pm$0.030 & 12.100$\pm$0.030 &        ...       &	     ...       &	...	  &	   ...     \\
RX~J0507.3$-$0326       & 11.454$\pm$0.021 & 10.780$\pm$0.024 & 10.628$\pm$0.022 & 10.508$\pm$0.023 & 10.478$\pm$0.019 & 10.599$\pm$0.090 & 8.989$^b$\\
TTS~J050717.9$-$032433  & 13.099$\pm$0.021 & 12.426$\pm$0.028 & 12.178$\pm$0.028 & 12.112$\pm$0.024 & 12.023$\pm$0.023 & 12.369$\pm$0.412 & 9.182$^b$\\
RX~J0507.4$-$0320       & 12.305$\pm$0.024 & 11.716$\pm$0.025 & 11.440$\pm$0.023 & 11.305$\pm$0.023 & 11.127$\pm$0.022 & 11.021$\pm$0.114 & 8.968$^b$\\
RX~J0507.4$-$0317       & 13.173$\pm$0.026 & 12.469$\pm$0.027 & 12.254$\pm$0.029 & 12.161$\pm$0.023 & 12.019$\pm$0.024 & 11.943$\pm$0.285 & 8.956$^b$\\
TTS~J050729.8$-$031705  & 14.937$\pm$0.035 & 14.230$\pm$0.040 & 13.964$\pm$0.053 & 13.821$\pm$0.029 & 13.545$\pm$0.038 &       12.570$^b$ & 8.995$^b$\\
TTS~J050730.9$-$031846  & 16.379$\pm$0.104 & 15.954$\pm$0.178 & 15.294$\pm$0.157 & 14.830$\pm$0.038 & 14.113$\pm$0.051 &       12.272$^b$ & 9.153$^b$\\
TTS~J050733.6$-$032517  & 14.707$\pm$0.040 & 14.138$\pm$0.044 & 13.839$\pm$0.055 & 13.667$\pm$0.028 & 13.470$\pm$0.036 &       12.457$^b$ & 9.030$^b$\\
TTS~J050734.8$-$031521  & 14.430$\pm$0.030 & 13.820$\pm$0.030 & 13.515$\pm$0.045 & 13.396$\pm$0.026 & 13.163$\pm$0.032 &       12.397$^b$ & 8.930$^b$\\
RX~J0507.6$-$0318       & 12.288$\pm$0.021 & 11.616$\pm$0.025 & 11.461$\pm$0.025 & 11.333$\pm$0.023 & 11.322$\pm$0.022 & 11.959$\pm$0.346 & 8.693$^b$\\
 TTS~J050741.0$-$032253 & 13.517$\pm$0.026 & 12.809$\pm$0.028 & 12.572$\pm$0.034 & 12.520$\pm$0.023 & 12.377$\pm$0.026 &       12.171$^b$ & 8.903$^b$\\
 TTS~J050741.4$-$031507 & 13.506$\pm$0.026 & 12.886$\pm$0.030 & 12.638$\pm$0.029 & 12.456$\pm$0.024 & 12.263$\pm$0.025 &       12.115$^b$ & 9.039$^b$\\
 TTS~J050752.0$-$032003 & 14.521$\pm$0.033 & 13.979$\pm$0.038 & 13.511$\pm$0.037 & 13.326$\pm$0.027 & 12.920$\pm$0.030 & 11.811$\pm$0.258 & 8.768$^b$\\
 TTS~J050801.4$-$032255 & 12.511$\pm$0.022 & 11.628$\pm$0.024 & 11.231$\pm$0.023 & 10.683$\pm$0.023 & 10.143$\pm$0.021 &  7.022$\pm$0.016 & 4.729$\pm$0.028\\
 TTS~J050801.9$-$031732 & 12.392$\pm$0.024 & 11.663$\pm$0.023 & 11.327$\pm$0.025 & 10.624$\pm$0.024 & 10.268$\pm$0.021 &  9.397$\pm$0.038 & 6.945$\pm$0.077\\
 TTS~J050804.0$-$034052 & 12.996$\pm$0.027 & 12.332$\pm$0.034 & 12.071$\pm$0.024 & 11.940$\pm$0.035 & 11.824$\pm$0.029 & 11.446$\pm$0.174 & 8.561$^b$\\
 TTS~J050836.6$-$030341 & 11.931$\pm$0.024 & 11.159$\pm$0.025 & 10.726$\pm$0.022 & 10.153$\pm$0.023 &  9.685$\pm$0.020 &  8.104$\pm$0.019 & 6.351$\pm$0.053\\
 TTS~J050845.1$-$031653 & 12.370$\pm$0.035 & 11.822$\pm$0.037 & 11.462$\pm$0.032 & 11.222$\pm$0.031 & 11.033$\pm$0.031 & 10.644$\pm$0.097 & 8.848$^b$\\
      RX~J0509.0$-$0315 &  9.914$\pm$0.024 &  9.530$\pm$0.024 &  9.408$\pm$0.021 &  9.337$\pm$0.023 &  9.360$\pm$0.019 &  9.321$\pm$0.035 & 8.706$^b$\\
      RX~J0510.1$-$0427 &  9.684$\pm$0.024 &  9.144$\pm$0.025 &  8.991$\pm$0.021 &  8.913$\pm$0.022 &  8.930$\pm$0.022 &  8.899$\pm$0.026 & 8.800$\pm$0.364\\
1RXS~J051011.5$-$025355 & 10.454$\pm$0.022 &  9.952$\pm$0.024 &  9.730$\pm$0.023 &  9.794$\pm$0.023 &  9.342$\pm$0.019 &  5.371$\pm$0.016 & 3.140$\pm$0.021\\
      RX~J0510.3$-$0330 &      10.038$^a$  &  9.806$\pm$0.060 &  9.749$\pm$0.049 &  9.177$\pm$0.022 &  9.205$\pm$0.019 &  9.283$\pm$0.032 & 8.941$\pm$0.409\\
1RXS~J051043.2$-$031627 & 10.079$\pm$0.026 &  9.735$\pm$0.022 &  9.648$\pm$0.025 &  9.558$\pm$0.022 &  9.584$\pm$0.020 &  9.534$\pm$0.036 & 8.555$^b$\\
      RX~J0511.7$-$0348 & 10.342$\pm$0.026 &  9.868$\pm$0.022 &  9.806$\pm$0.023 &  9.700$\pm$0.023 &  9.702$\pm$0.019 &  9.659$\pm$0.041 & 8.537$^b$\\
      RX~J0512.3$-$0255 & 10.425$\pm$0.023 &  9.688$\pm$0.023 &  9.140$\pm$0.019 &  8.426$\pm$0.023 &  8.043$\pm$0.020 &  7.288$\pm$0.016 & 4.479$\pm$0.027\\
\hline\\
\end{longtable}
\footnotesize{Notes: $^a$ Upper limit on magnitude. The source is not resolved in a consistent fashion with the other bands. $^b$ This magnitude corresponds to upper limit ($S/N<2$).}

\pagestyle{empty}
\setlength{\tabcolsep}{6pt}
\begin{landscape}
\scriptsize
\begin{longtable}{lrcrc|rcrcrc|cc}
\caption[ ]{\label{tab:ew_flux} Line equivalent widths and line fluxes for hydrogen and helium lines.}\\
\hline\hline
   &  & {\sc Hydrogen lines} & & &  & & & {\sc Helium lines} & &  & & \\
ID name  & $EW_{\rm H\alpha}$&$\log F^{\rm H\alpha}$&$EW_{\rm H\beta}$&$\log F^{\rm H\beta}$& $EW_{\rm \lambda5876}$ &$\log F^{\rm \lambda5876}$&$EW_{\rm \lambda6678}$&$\log F^{\rm \lambda6678}$&$EW_{\rm \lambda7065}$&$\log F^{\rm \lambda7065}$ & Accretor & Notes \\
	 & (\AA)             & (erg\,s$^{-1}$\,cm$^2$)       		& (\AA)	    & (erg\,s$^{-1}$\,cm$^2$)			  & (\AA)	    &	 (erg\,s$^{-1}$\,cm$^2$)	  & (\AA) 	  &  (erg\,s$^{-1}$\,cm$^2$)	      & (\AA) 	  &  (erg\,s$^{-1}$\,cm$^2$)  & (Y/N)&  \\
\hline
\endfirsthead
\caption{continued.}\\
\hline\hline
   &  & {\sc Hydrogen lines} & & &  & & & {\sc Helium lines} & &  & & \\
ID name  & $EW_{\rm H\alpha}$&$\log F^{\rm H\alpha}$&$EW_{\rm H\beta}$&$\log F^{\rm H\beta}$& $EW_{\rm \lambda5876}$ &$\log F^{\rm \lambda5876}$&$EW_{\rm \lambda6678}$&$\log F^{\rm \lambda6678}$&$EW_{\rm \lambda7065}$&$\log F^{\rm \lambda7065}$ & Accretor & Notes \\
	 & (\AA)             & (erg\,s$^{-1}$\,cm$^2$)       		& (\AA)	    & (erg\,s$^{-1}$\,cm$^2$)			  & (\AA)	    &	 (erg\,s$^{-1}$\,cm$^2$)	  & (\AA) 	  &  (erg\,s$^{-1}$\,cm$^2$)	      & (\AA) 	  &  (erg\,s$^{-1}$\,cm$^2$)  & (Y/N)&  \\
\hline
\endhead
\hline
\endfoot
\hline
1RXS~J045912.4$-$033711 &    1.80$\pm$0.20& ...&	      &    &		    &	 &		  &    &		&     & N& G  \\
1RXS~J050416.9$-$021426 &    0.03$\pm$0.20& ...&	      &    &		    &	 &		  &    &		&     & N& G  \\ 
 TTS~J050513.5$-$034248 & $-$9.80$\pm$0.40& 6.2&	      &    &		    &	 &		  &    &		&     & N& G  \\ 
 TTS~J050538.9$-$032626 & $-$4.90$\pm$0.20& 6.2&	      &    &		    &	 &		  &    &		&     & N& G \\ 
      RX~J0506.6$-$0337 &    1.00$\pm$0.20& ...&	      &    &		    &	 &		  &    &		&     & N& G \\ 
 TTS~J050644.4$-$032913 & $-$4.70$\pm$0.30& 6.0&	      &    &		    &	 &		  &    &		&     & N& G \\ 
TTS~J050646.1$-$031922  & $-$35.2$\pm$0.8 & 8.0&	      &    &		    &	 &   ...	  &... &  ...		& ... & Y &   \\
		        & $-$34.8$\pm$1.5 & 8.0&	      &    &		    &	 &   ...	  &... &  ...		& ... & &      \\
RX~J0506.8$-$0318       &  $-$1.3$\pm$0.1 & 6.1&	      &    &		    &	 &   ...	  &... &  ...		& ... & N&  \\
		        &  $-$1.2$\pm$0.1 & 6.1&	      &    &		    &	 &   ...	  &... &  ...		& ... & &     \\
		        &  $-$1.5$\pm$0.1 & 6.2&	      &    &		    &	 &   ...	  &... &  ...		& ... & &	\\
TTS~J050647.5$-$031910  &  $-$0.7$\pm$0.1 & 4.8&	      &    &		    &	 &   ...	  &... &  ...		& ... & N&   \\
		        &  $-$0.7$\pm$0.1 & 4.8&	      &    &		    &	 &   ...	  &... &  ...		& ... & &    \\
		        &  $-$0.3$\pm$0.1 & 4.4&	      &    &		    &	 &   ...	  &... &  ...		& ... & &    \\
RX~J0506.8$-$0327       &  $-$6.0$\pm$0.2 & 6.1&	      &    &		    &	 &   ...	  &... &  ...		& ... & N& \\
		        &  $-$5.8$\pm$0.1 & 6.1&	      &    &		    &	 &   ...	  &... &  ...		& ... & &	 \\
RX~J0506.8$-$0305       &  $-$4.7$\pm$0.1 & 5.8&	      &    &		    &	 &$-$0.25$\pm$0.01& 4.3&  ...		& ... & N&  \\
TTS~J050649.8$-$031933  & $-$15.5$\pm$1.00& 6.5&	      &    &		    &    &		  &    &		&     & Y& G \\
TTS~J050649.8$-$032104  & $-$65.7$\pm$0.6 & 7.4&	      &    &		    &	 &$-$0.98$\pm$0.06& 5.3&  ...		& ... & Y &  \\
		        & $-$25.4$\pm$0.3 & 7.0&	      &    &		    &	 &   ...	  &... &  ...		& ... & &      \\
TTS~J~050650.5$-$032014 &$-$14.00$\pm$1.00& 6.1&	      &    &		    &    &		  &    &		&     & N& G \\
TTS~J050650.7$-$032008  &$-$26.00$\pm$0.50& 6.6&	      &    &		    &	 &		  &    &		&     & Y& G \\
RX~J0506.9$-$0319~NW    &  $-$1.5$\pm$0.1 & 5.7&	      &    &		    &	 &   ...	  &... &  ...		& ... & N&  \\
RX~J0506.9$-$0319~SE    &  $-$4.1$\pm$0.2 & 6.9&	      &    &		    &	 &   ...	  &... &  ...		& ... & Y&  \\
		        &  $-$4.2$\pm$0.1 & 6.9&	      &    &		    &	 &   ...	  &... &  ...		& ... & &     \\
RX~J0506.9$-$0320~W     &  $-$2.1$\pm$0.1 & 6.4&	      &    &		    &	 &   ...	  &... &  ...		& ... & N&   \\
RX~J0506.9$-$0320~E     &  $-$1.4$\pm$0.1 & 6.8&	      &    &		    &	 &   ...	  &... &  ...		& ... & N&$a$ \\
		        &  $-$1.6$\pm$0.1 & 6.9&	      &    &		    &	 &   ...	  &... &  ...		& ... & &     \\
		        &  $-$1.6$\pm$0.1 & 6.9&	      &    &		    &	 &   ...	  &... &  ...		& ... & &     \\
		        &  $-$1.8$\pm$0.1 & 6.9&	      &    &		    &	 &   ...	  &... &  ...		& ... & &      \\
TTS~J050654.5$-$032046  &$-$60.00$\pm$5.00& 7.3&	      &    &		    &	 &		  &    &		&     & Y&  G \\  
LkH$\alpha$~333         & $-$48.0$\pm$0.8 & 8.1&	      &    &		    &	 &$-$0.08$\pm$0.01& 5.4&$-$0.14$\pm$0.01& 5.6 & Y&  \\
		        & $-$47.2$\pm$3.4 & 8.1&$-$8.6$\pm$0.2& 7.4&$-$0.37$\pm$0.01& 6.1&$-$0.02$\pm$0.01& 4.8&		&     &  &   \\
		        &	 ...	  & ...&$-$8.9$\pm$0.3& 7.5&	    ...     & ...&   ...	  & ...&		&     &  &	\\
		        & $-$53.4$\pm$1.4 & 8.2&$-$8.8$\pm$0.2& 7.5&$-$0.51$\pm$0.01& 6.2&$-$0.04$\pm$0.01& 5.0&		&     &  &  \\
		        & $-$55.2$\pm$1.8 & 8.2&$-$9.5$\pm$0.1& 7.5&$-$0.71$\pm$0.02& 6.4&$-$0.10$\pm$0.01& 5.4&		&     &  & \\
		        & $-$51.9$\pm$1.6 & 8.2&$-$9.1$\pm$0.5& 7.5&$-$0.55$\pm$0.01& 6.2&$-$0.08$\pm$0.01& 5.4&		&     &  & \\
		        & $-$57.9$\pm$0.2 & 8.2&$-$9.6$\pm$0.3& 7.5&$-$0.62$\pm$0.01& 6.3&$-$0.11$\pm$0.01& 5.5&		&     &  &	\\
L1616~MIR~4	        & $-$28.6$\pm$0.3 & 8.1&	      &    &		    &	 &   ...	  &... &  ...		& ... & Y&  \\      
		        & $-$25.2$\pm$0.6 & 8.0&	      &    &		    &	 &   ...	  &... &  ...		& ... & &      \\   
		        & $-$26.5$\pm$0.3 & 8.0&	      &    &		    &	 &   ...	  &... &  ...		& ... & &      \\   
		        & $-$26.0$\pm$0.8 & 8.0&	      &    &		    &	 &   ...	  &... &  ...		& ... & &	\\  
RX~J0507.0$-$0318       &  $-$1.3$\pm$0.1 & 5.9&	      &    &		    &	 &   ...	  &... &  ...		& ... & N&  \\	    
		        &  $-$1.0$\pm$0.1 & 5.8&	      &    &		    &	 &   ...	  &... &  ...		& ... & &	\\  
		        &  $-$1.4$\pm$0.1 & 6.0&	      &    &		    &	 &   ...	  &... &  ...		& ... & &     \\    
TTS~J050657.0$-$031640  &$-$94.00$\pm$2.00& 7.2&	      &    &		    &	 &		  &    &		&     & Y& G \\   
TTS~J050704.7$-$030241  &$-$17.50$\pm$1.00& 6.3&	      &    &		    &	 &		  &    &		&     & N& G \\   
TTS~J050705.3$-$030006  & $-$2.00$\pm$0.40& 6.2&	      &    &		    &	 &		  &    &		&     & N& G \\   
RX~J0507.1$-$0321       & $-$37.1$\pm$0.9 & 7.3&	      &    &		    &	 &$-$0.53$\pm$0.01& 5.2&$-$0.37$\pm$0.01& 5.3 & Y& \\   
		        & $-$36.9$\pm$0.5 & 7.3&	      &    &		    &	 &$-$0.51$\pm$0.01& 5.2&$-$0.40$\pm$0.01& 5.3 & &   \\ 
TTS~J050706.2$-$031703  &$-$13.50$\pm$0.50& 6.1&	      &    &		    &    &		  &    &		&     & N& G \\  
RX~J0507.2$-$0323       &  $-$1.2$\pm$0.1 & 6.5&	      &    &		    &	 &   ...	  &... &  ...		& ... & N&    \\   
TTS~J050713.5$-$031722  &$-$10.50$\pm$0.50& 7.1&	      &    &		    &    &		  &    &		&     & N& G \\  
RX~J0507.3$-$0326       &  $-$2.1$\pm$0.1 & 6.1&	      &    &		    &	 &   ...	  &... &  ...		& ... & N&   \\      
		        &  $-$1.8$\pm$0.1 & 6.1&	      &    &		    &	 &   ...	  &... &  ...		& ... & &	\\  
TTS~J050717.9$-$032433  &  $-$3.4$\pm$0.1 & 6.1&	      &    &		    &	 &   ...	  &... &  ...		& ... & N&  \\	    
		        &  $-$4.0$\pm$0.1 & 6.1&	      &    &		    &	 &$-$0.11$\pm$0.01& 4.4&  ...		& ... & &      \\   
RX~J0507.4$-$0320       &  $-$4.3$\pm$0.1 & 5.8&	      &    &		    &	 &$-$0.16$\pm$0.02& 4.2&  ...		& ... & N&  \\	    
		        &  $-$3.8$\pm$0.1 & 5.8&	      &    &		    &	 &$-$0.08$\pm$0.01& 3.9&  ...		& ... & &    \\     
		        &  $-$3.9$\pm$0.1 & 5.8&	      &    &		    &	 &$-$0.09$\pm$0.01& 3.9&  ...		& ... & &    \\     
RX~J0507.4$-$0317       &  $-$2.7$\pm$0.1 & 6.0&	      &    &		    &	 &   ...	  &... &  ...		& ... & N&   \\      
		        &  $-$6.0$\pm$0.1 & 6.3&	      &    &		    &	 &$-$0.09$\pm$0.01& 4.3&  ...		& ... & &    \\     
		        &  $-$5.7$\pm$0.1 & 6.3&	      &    &		    &	 &$-$0.16$\pm$0.01& 4.5&  ...		& ... & &     \\    
TTS~J050729.8$-$031705  &  $-$4.8$\pm$0.1 & 5.6&	      &    &		    &	 &   ...	  &... &  ...		& ... & N&   \\    
		        &  $-$0.8$\pm$0.1 & 4.9&	      &    &		    &	 &   ...	  &... &  ...		& ... & &    \\   
		        &  $-$9.2$\pm$0.2 & 5.9&	      &    &		    &	 &   ...	  &... &  ...		& ... & &    \\   
TTS~J050730.9$-$031846  & $-$52.0$\pm$1.4 & 6.9&	      &    &		    &	 &   ...	  &... &  ...		& ... & Y&    \\   
		        & $-$69.3$\pm$2.6 & 7.1&	      &    &		    &	 &   ...	  &... &  ...		& ... & &     \\  
TTS~J050733.6$-$032517  & $-$2.30$\pm$0.20& 5.5&	      &    &		    &	 &		  &    &		&     & N& G \\   
TTS~J050734.8$-$031521  &  $-$1.6$\pm$0.1 & 5.5&	      &    &		    &	 &   ...	  &... &  ...		& ... & N&  \\	  
		        &  $-$2.7$\pm$0.1 & 5.7&	      &    &		    &	 &   ...	  &... &  ...		& ... & &      \\ 
		        &  $-$4.6$\pm$0.1 & 5.9&	      &    &		    &	 &$-$2.24$\pm$0.07& 5.4&  ...		& ... & &$b$\\ 
RX~J0507.6$-$0318       &  $-$1.2$\pm$0.1 & 6.1&	      &    &		    &	 &   ...	  &... &  ...		& ... & N&   \\    
		        &  $-$1.6$\pm$0.1 & 6.2&	      &    &		    &	 &   ...	  &... &  ...		& ... & &   \\    
		        &  $-$1.6$\pm$0.1 & 6.2&	      &    &		    &	 &   ...	  &... &  ...		& ... & &    \\   
 TTS~J050741.0$-$032253 & $-$4.30$\pm$0.10& 5.9&	      &    &		    &	 &		  &    &		&     & N& G \\ 
 TTS~J050741.4$-$031507 & $-$6.80$\pm$0.40& 6.1&	      &    &		    &	 &		  &    &		&     & N& G \\ 
 TTS~J050752.0$-$032003 & $-$9.50$\pm$0.20& 6.1&	      &    &		    &	 &		  &    &		&     & N& G \\ 
 TTS~J050801.4$-$032255 &$-$21.50$\pm$0.50& 7.2&	      &    &		    &	 &		  &    &		&     & Y& G \\ 
 TTS~J050801.9$-$031732 &$-$15.50$\pm$0.50& 6.9&	      &    &		    &	 &		  &    &		&     & Y& G \\ 
 TTS~J050804.0$-$034052 & $-$2.20$\pm$0.20& 5.9&	      &    &		    &	 &		  &    &		&     & N& G \\ 
 TTS~J050836.6$-$030341 &$-$68.50$\pm$1.00& 7.4&	      &    &		    &	 &		  &    &		&     & Y& G \\ 
 TTS~J050845.1$-$031653 & $-$6.70$\pm$1.00& 6.1&	      &    &		    &	 &		  &    &		&     & N& G \\ 
      RX~J0509.0$-$0315 &    1.10$\pm$0.20& ...&	      &    &		    &	 &		  &    &		&     & N& G \\ 
      RX~J0510.1$-$0427 & $-$0.20$\pm$0.20& 5.8&	      &    &		    &	 &		  &    &		&     & N& G \\ 
1RXS~J051011.5$-$025355 & $-$0.10$\pm$0.20& 5.7&	      &    &		    &	 &		  &    &		&     & N& G \\ 
      RX~J0510.3$-$0330 &    1.50$\pm$0.20& ...&	      &    &		    &	 &		  &    &		&     & N& G \\ 
1RXS~J051043.2$-$031627 &    2.40$\pm$0.20& ...&	      &    &		    &	 &		  &    &		&     & N& G \\ 
      RX~J0511.7$-$0348 &    1.80$\pm$0.20& ...&	      &    &		    &	 &		  &    &		&     & N& G \\ 
      RX~J0512.3$-$0255 & $-$6.50$\pm$0.50& 7.4&	      &    &		    &	 &		  &    &		&     & Y& G \\ 
\hline							  
\end{longtable}
\footnotesize{Notes:
The columns list: star name (column 1); equivalent widths and observed fluxes of the H$\alpha$, H$\beta$, \ion{He}{i} $\lambda$5876 \AA, 
$\lambda$6678 \AA, and $\lambda$7065 \AA\ lines (columns 2--11); our classification as accretor or not (column 12); some notes (G: 
$EW_{\rm H\alpha}$ measured by \citealt{gandolfietal2008}; $a$: the $EW_{\rm H\alpha}$ refers to the emission in the central reversal; 
$b$: the helium emission is probably due to flare presence).
}
\end{landscape}
\normalsize

\begin{table}
\caption[ ]{\label{tab:accret_param} Mean accretion luminosities, stellar masses, and mass accretion rates derived 
for the 15 accretors in L1615/L1616 from the Ba98+Ch00, DM97, and PS99 stellar masses.}
\begin{center}
\begin{tabular}{l|c|cc|cc|cc}
\hline\hline
ID name & $\log \langle\frac{L_{\rm acc}}{L_\odot}\rangle$ &  \multicolumn{2}{c|}{\rm Ba98+Ch00 } &  \multicolumn{2}{c|}{\rm DM97 }  & \multicolumn{2}{c}{\rm PS99 }  \\
         &                                                 & $M_\star$ & $\log \dot M_{\rm acc}$ & $M_\star$ & $\log \dot M_{\rm acc}$ & $M_\star$ & $\log \dot M_{\rm acc}$\\
	 &        & ($M_\odot$) & ($M_\odot {\rm \,yr^{-1}}$) & ($M_\odot$) & ($M_\odot {\rm \,yr^{-1}}$) & ($M_\odot$) & ($M_\odot {\rm \,yr^{-1}}$) \\
\hline
TTS~J050646.1$-$031922 & $-$1.0 & ...  &   ...   & 0.75 &  $-$8.0 & 1.35 &  $-$8.3 \\
TTS~J050649.8$-$031933 & $-$2.6 & 0.47 &  $-$9.4 & 0.22 &  $-$9.1 & 0.25 &  $-$9.1 \\
TTS~J050649.8$-$032104 & $-$1.6 & 0.87 &  $-$8.5 & 0.35 &  $-$8.1 & 0.50 &  $-$8.3 \\
TTS~J050650.7$-$032008 & $-$2.9 & 0.26 &  $-$9.6 & 0.19 &  $-$9.5 & 0.18 &  $-$9.5 \\
RX~J0506.9$-$0319~SE   & $-$1.9 & 1.40 &  $-$9.0 & 0.45 &  $-$8.5 & 0.93 &  $-$8.8 \\
TTS~J050654.5$-$032046 & $-$3.0 & 0.26 & $-$10.0 & 0.25 & $-$10.0 & 0.22 & $-$10.0 \\
LkH$\alpha$~333        & $-$0.3 & ...  &   ...   & 0.75 &  $-$7.0 & 1.70 &  $-$7.3 \\
L1616~MIR~4	       & $-$0.7 & ...  &   ...   & 1.70 &  $-$7.9 & 1.75 &  $-$8.0 \\ 
TTS~J050657.0$-$031640 & $-$2.3 & 0.27 &  $-$9.0 & 0.19 &  $-$8.9 & 0.18 &  $-$8.8 \\
RX~J0507.1$-$0321      & $-$1.8 & 0.80 &  $-$8.9 & 0.40 &  $-$8.6 & 0.50 &  $-$8.7 \\
TTS~J050730.9$-$031846 & $-$3.8 & 0.11 & $-$10.7 & 0.15 & $-$10.9 & 0.13 & $-$10.8 \\ 
TTS~J050801.4$-$032255 & $-$2.2 & 0.88 &  $-$9.4 & 0.47 &  $-$9.1 & 0.56 &  $-$9.2 \\
TTS~J050801.9$-$031732 & $-$2.5 & 0.80 &  $-$9.6 & 0.44 &  $-$9.4 & 0.50 &  $-$9.4 \\
TTS~J050836.6$-$030341 & $-$1.6 & 0.75 &  $-$8.6 & 0.32 &  $-$8.2 & 0.44 &  $-$8.4 \\
RX~J0512.3$-$0255      & $-$1.6 & ...  &   ...   & 1.30 &  $-$8.9 & 1.50 &  $-$8.9 \\
\hline							  
\end{tabular}
\end{center}
\end{table}

\topmargin -1 cm
\twocolumn
\begin{appendix}

\section{Radial velocity and lithium abundance determinations} 
\label{sec:RV_NLi}

\subsection{Radial velocity}
\label{sec:rad_vel}
We were able to measure radial velocities for 19 objects out of the 23 YSOs
for which FLAMES spectroscopy was available. We followed the same 
procedure as in \cite{biazzoetal2012}. Heliocentric RVs of the targets were determined 
through the task {\sc fxcor} within the IRAF package {\sc rv}, which cross-correlates  
the target and template spectra, excluding regions affected by broad lines
or prominent telluric features. As GIRAFFE templates, we used RX~J0507.2$-$0323 and 
RX~J0507.0$-$0318 for the earliest-type and latest-type stars, respectively, while as UVES 
template we considered the first spectrum acquired for LkH$\alpha$~333.
We measured the RV of each template using the IRAF task {\sc rvidlines} inside the {\sc rv} 
package. This task measures RVs from a line list. We used 40 and 10 lines for 
the  UVES and GIRAFFE spectra, respectively, obtaining 
$V_{\rm rad}=24.8 \pm 1.7$ km s$^{-1}$ for RX~J0507.2$-$0323, 
$V_{\rm rad}=23.7 \pm 1.9$ km s$^{-1}$ for RX~J0507.0$-$0318, and 
$V_{\rm rad}=27.0 \pm 0.5$ km s$^{-1}$ for LkH$\alpha$~333.  

The centroids of the cross-correlation  function (CCF) peaks were determined by adopting 
Gaussian fits, and the RV errors were computed by {\sc fxcor} according to the fitted 
peak  height and the antisymmetric noise (see \citealt{tonrydavis1979}). 
The RV values derived for each spectrum are listed in Table~\ref{tab:all_param}  
with the corresponding uncertainties.

\begin{figure}[h!]
\begin{center}
 \begin{tabular}{c}
\hspace{-.6cm}
\includegraphics[width=9.5cm]{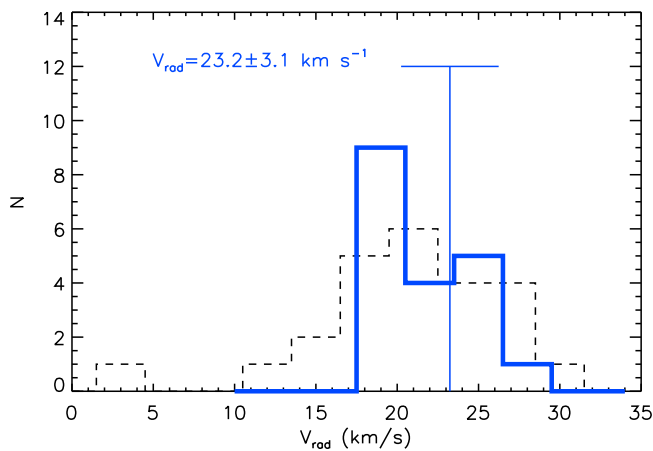}
\vspace{-.4cm}
 \end{tabular}
\caption{Average RV distribution of the L1615/L1616 low-mass stars (thick solid line). 
The bar indicates the mean RV value and its standard deviation from the average. 
The dashed line represents the distribution obtained by \cite{alcalaetal2004} 
for 24 stars in L1616.
}
\label{fig:vrad_distr}
 \end{center}
\end{figure}

In Fig.~\ref{fig:vrad_distr}, we show the L1615/L1616 distribution of the RV measurements 
obtained from both the UVES and GIRAFFE spectra. When more than one spectrum was acquired, we 
computed the average RV for each object. 

\subsection{Lithium abundance}
\label{sec:lithium_abundance}
Lithium equivalent widths ($EW_{\rm Li}$) were measured by direct integration or 
by Gaussian fit using the IRAF task {\sc splot}. Errors in 
$EW_{\rm Li}$ were estimated in the following way: $i)$ when only one spectrum 
was available, the standard deviation of three $EW_{\rm Li}$ measurements was 
adopted; $ii)$ when more than one spectrum was gathered, the standard deviation 
of the measurements on the different spectra was adopted. Typical errors in 
$EW_{\rm Li}$ are of 1--20 m\AA. Our $EW_{\rm Li}$ measurements are consistent 
with the values of \cite{gandolfietal2008} within 17\,m\AA\,on the average.

Mean lithium abundances ($\log n{\rm (Li)}$) were estimated from the average 
$EW_{\rm Li}$ listed in Table~\ref{tab:all_param} and the effective temperature 
($T_{\rm eff}$) from \cite{gandolfietal2008}, by using the LTE curves-of-growth 
reported by \cite{pavlenkomagazzu1996} for $T_{\rm eff}>4000$~K, and by 
\cite{pallaetal2007} for $T_{\rm eff}<4000$~K. The $\log g$ values were 
derived as explained in Sect.~\ref{sec:accretion_rates}. 
The main source of error in $\log n{\rm (Li)}$ comes from the uncertainty 
in $T_{\rm eff}$, which is $\Delta T_{\rm eff}\sim100$~K (\citealt{gandolfietal2008}). 
Taking this value and a mean error of 10\,m\AA\ in $EW_{\rm Li}$ into account, 
we estimated an uncertainty in $\log n{\rm (Li)}$ ranging from $\sim$0.05$-$0.30 dex 
for cooler stars ($T_{\rm eff}\sim 3000$ K) down to $\sim$0.05$-$0.15 dex for 
warmer stars ($T_{\rm eff}\sim 5000$ K), depending on the $EW_{\rm Li}$ value. 
Moreover, the $\log g$ value affects the lithium abundance, in the sense that 
the lower the surface gravity the higher the lithium abundance, and vice versa. 
In particular, the difference in $\log n{\rm (Li)}$ may rise to $\sim \pm0.1$ dex 
when considering stars with mean values of $EW_{\rm Li}=500$ (m\AA)\, and 
$T_{\rm eff}=4500$ K and assuming $\Delta \log g = \mp 0.15$ dex, which is the 
mean error derived from the \cite{gandolfietal2008} stellar parameters.

\section{$M_{\star}$ and $\dot M_{\rm acc}$ as determined from different evolutionary tracks}
\label{sec:Mass_Macc_tracks}

\cite{gandolfietal2008} have derived the mass of 56 members of L1615/L1616 by comparing 
the location of each object on the HR diagram with the theoretical PMS evolutionary 
tracks by \cite{baraffeetal1998} and \cite{chabrieretal2000} - Ba98+Ch00, 
\cite{dantonamazzitelli1997} - DM97, and \cite{pallastahler1999} - PS99, which are 
available in the mass ranges $0.003 M_\odot \le M \le 1.40 M_\odot$, 
$0.017 M_\odot \le M \le 3 M_\odot$, $0.1 M_\odot \le M \le 6 M_\odot$, for Ba98+Ch00, DM97, and PS99 
models, respectively. 
The usage of masses computed from different evolutionary tracks allowed us to estimate the 
model-dependent uncertainties on $\dot M_{\rm acc}$ associated with the derived 
masses. 

\begin{figure*}	
\begin{center}
 \begin{tabular}{c}
\hspace{-.6cm}
\includegraphics[width=16cm]{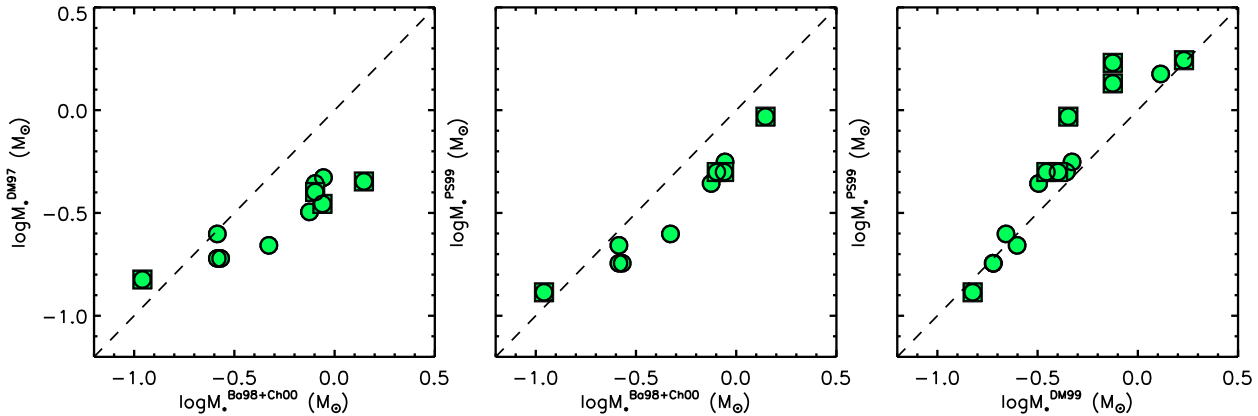}
\vspace{-.2cm}
 \end{tabular}
\caption{Comparison between masses derived from the Ba98+Ch00 and DM97 tracks ({\it left panel}), 
the Ba98+Ch00 and PS99 tracks ({\it middle panel}), and the DM97 and PS99 tracks ({\it right panel}) 
evolutionary tracks. Symbols are as in Fig.~\ref{fig:2MASS_color_color}.}
\label{fig:mass_tracks}
 \end{center}
\end{figure*}

\begin{figure*}	
\begin{center}
 \begin{tabular}{c}
\hspace{-.6cm}
\includegraphics[width=16cm]{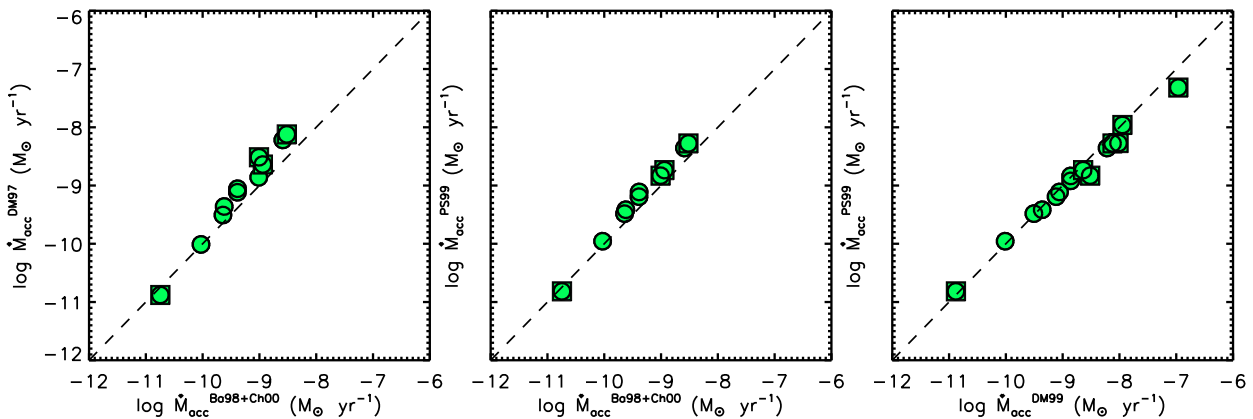}
\vspace{-.2cm}
 \end{tabular}
\caption{Comparison between mass accretion rates derived using the  Ba98+Ch00 and DM97 tracks({\it left panel}), 
the Ba98+Ch00 and PS99 tracks ({\it middle panel}), and the DM97 and PS99 tracks ({\it right panel}). 
Symbols are as in Fig.~\ref{fig:2MASS_color_color}. }
\label{fig:macc_tracks}
 \end{center}
\end{figure*}

In Fig.~\ref{fig:mass_tracks}, we show the comparison between the masses derived from 
the three sets of tracks for the accreting objects. The largest residuals are seen when comparing the Ba98+Ch00 
and DM97 tracks. Fig.~\ref{fig:macc_tracks} shows the comparisons of the mass accretion 
rates when calculated using the different evolutionary tracks. It can be seen that the 
$\log \dot M_{\rm acc}$ values are rather independent of the choice of the PMS track. 
Therefore, it is the model-dependent uncertainty on mass which produces the dispersion 
in the $\log \dot M_{\rm acc}$ vs. $\log M_\star$ plot, whereas the $\log \dot M_{\rm acc}$ 
values are practically insensitive to the choice of the evolutionary model.

\section{Notes on individual objects}
\label{sec:notes_on_ind_objects}

\subsection{TTS~050730.9$-$031846: a sub-luminous YSO with edge-on disk?}
\label{sec:sub-luminous}

The star TTS~050730.9$-$031846 appears to be sub-luminous in the HR diagram 
reported by \cite{gandolfietal2008}, in comparison with typical YSOs in L1615/L1616 
of similar effective temperature. Sub-luminous YSOs have been found in other SFRs, 
like L1630N, L1641, Lupus, and Taurus (see \citealt{fangetal2009, fangetal2013, 
comeronetal2003, whitehillenbrand2004, looperetal10, alcalaetal2014}). 
Among the hypotheses to explain the sub-luminosity of these YSOs are the following: 
$i)$~These YSOs are believed to be embedded or to harbor flared disks with high 
inclination angles. In this case, the stellar photospheric light and any emission 
due to physical processes in the inner disk region are severely suppressed by the 
edge-on disk; $ii)$~Other authors (\citealt{baraffechabrier2010}) argue that episodic 
strong accretion during the PMS evolution produces objects with smaller radius, 
higher central temperatures, and hence lower luminosity, compared to the 
non-accreting counterparts of the same age and mass.

The anomalous position of TTS~050730.9$-$031846 on the $J-H$ versus $H-K{\rm s}$ diagram 
(Fig.~\ref{fig:2MASS_color_color}) would favor the hypothesis
of a YSO with a high inclination angle in which both the stellar luminosity
and the accretion luminosity are suppressed by the optically thick edge-on disk. 
A star with the effective temperature of TTS~050730.9$-$031846 ($\sim 3100$ K), at the 
age of L1615/L1616 ($\sim 3$ Myr; \citealt{gandolfietal2008}) should have a mass 
$M_\star \sim 0.1\,M_{\odot}$ and should be a factor $\sim 5$ more luminous than observed. 
This would imply that the mass accretion rate of the star should be a factor (5)$^{1.5}$ 
higher than observed (\citealt{alcalaetal2014}). This means
values of $\sim 1.8, 1.3, 1.4\times10^{-10}$ $M_\odot$\,yr$^{-1}$ 
for the masses drawn from the Ba98+Ch00, DM97, PS99 tracks, respectively, 
i.e. similar to the $\dot M_{\rm acc}$ measured in YSOs with the same 
mass (see Fig.~\ref{fig:Macc_Mass_tracks}). 

An edge-on disk may also produce variable circumstellar extinction, 
inducing changes in the continuum that produce strong variations of 
the equivalent width of emission lines (see, e.g., the case of the transitional object 
T~Chamaeleontis studied by \citealt{schisanoetal2009}). 
In fact, the $EW_{\rm H\alpha}=-290.00\pm30.00$\,\AA\ of 
TTS~050730.9$-$031846 measured by \cite{gandolfietal2008} is a factor of 
about six higher than what we measured here. Whether such strong 
variability is due to variable circumstellar extinction is not clear, 
but can only be confirmed by a simultaneous multi-band photometric 
monitoring of the star.

\subsection{TTS~050649.8$-$031933}
\label{TTS050649.8-031933}
This star was classified as a WTTs by \cite{gandolfietal2008}, but re-analyzing the three 
spectra acquired by the authors, helium and oxygen lines always appear in emission (as also 
reported in their Table 4). The results of our three measurements are 
$EW_{\rm H\alpha}=-15.5, -21.7, -17.6$ \AA. Moreover, the $WISE$ colors of the star are consistent 
with those of a Class II YSO. We thus classify TTS~050649.8$-$031933 as an accreting YSO.

\subsection{TTS~050649.8$-$032104 and L1616~MIR~4}
\label{sec:TTS050649.8-032104}
Besides the sub-luminous object discussed above, these are the other two YSOs in our 
sample with the strongest variations in the H$\alpha$ line (see Fig.~\ref{fig:ew_comparison}). 
In particular, for TTS~050649.8$-$032104 \cite{gandolfietal2008} measured 
$EW_{\rm H\alpha}=-195\pm5$\,\AA, i.e. about three times higher than what we measure here, implying a 
difference of $\sim 0.6$\,dex in $\log \dot M_{\rm acc}$. In the case of L1616~MIR~4, the 
\cite{gandolfietal2008} result ($EW_{\rm H\alpha}=-60\pm6$ \AA), is also about three times
higher than our measurements, corresponding to a difference of 
$\sim 0.4$\,dex in $\log \dot M_{\rm acc}$. 

\subsection{TTS~050713.5$-$031722}
This star was classified as a CTTs by \cite{gandolfietal2008}, 
but re-analyzing their spectrum (see their Table~4), neither helium 
nor oxygen lines appear in emission (as reported by the authors) and we measured an H$\alpha$ 
equivalent width of $\sim -8$~\AA. Considering its spectral type (K8.5), it 
can be classified as non-accretor according to the \cite{whitebasri2003} criteria.

\subsection{Other targets}
\label{othertargets}

In some of the $WISE/2MASS$ color-color diagrams of Fig.~\ref{fig:WISE_color_color}, the targets 
TTS~050647.5$-$031910, TTS~050706.2$-$031703, TTS~050752.0$-$032003, 1RXS~J051011.5$-$025355, 
TTS~050734.8$-$031521, and TTS~050729.8$-$031705 appear to be close to or within the regions of 
Class~II objects, according to \cite{koenigetal2012}. We indeed classified these 6 objects as 
non-accretors (see Table~\ref{tab:ew_flux}), because their H$\alpha$ equivalent widths and spectral 
types are consistent with this object class, according to the \cite{whitebasri2003} criteria.

\end{appendix}


\end{document}